\documentclass[%
superscriptaddress,
preprint,
 amsmath,amssymb,
 aps,
]{revtex4-2}
\usepackage{svg}
\usepackage{booktabs}
\usepackage{longtable}
\usepackage{csvsimple}
\usepackage{pgfplots}
\usepackage[version=4]{mhchem}
\usepackage{siunitx}
\usepackage{tikz}
\usepackage{xcolor}
\usetikzlibrary{quantikz2}
\pgfplotsset{compat=1.17}

\definecolor{dualAnalagous1}{HTML}{06386C}
\definecolor{dualAnalagous1a}{HTML}{043F40}
\definecolor{dualAnalagous1b}{HTML}{0D0BBC}
\definecolor{dualAnalagous2}{HTML}{DF780D}
\definecolor{dualAnalagous2a}{HTML}{F66361}
\definecolor{dualAnalagous2b}{HTML}{979909}
\definecolor{neonorange}{HTML}{FF5C00}

\definecolor{triad1}{HTML}{06386C}
\definecolor{triad2}{HTML}{914B45}
\definecolor{triad3}{HTML}{879145}

\definecolor{mpltab1}{rgb}{0.12156862745098039, 0.4666666666666667, 0.7058823529411765}
\definecolor{mpltab2}{rgb}{1.0, 0.4980392156862745, 0.054901960784313725}
\definecolor{mpltab3}{rgb}{0.17254901960784313, 0.6274509803921569, 0.17254901960784313}
\definecolor{mpltab4}{rgb}{0.8392156862745098, 0.15294117647058825, 0.1568627450980392}
\definecolor{mpltab5}{rgb}{0.5803921568627451, 0.403921568627451, 0.7411764705882353}
\definecolor{mpltab6}{rgb}{0.5490196078431373, 0.33725490196078434, 0.29411764705882354}
\definecolor{mpltab7}{rgb}{0.8901960784313725, 0.4666666666666667, 0.7607843137254902}
\definecolor{mpltab8}{rgb}{0.4980392156862745, 0.4980392156862745, 0.4980392156862745}
\definecolor{mpltab9}{rgb}{0.7372549019607844, 0.7411764705882353, 0.13333333333333333}
\definecolor{mpltab10}{rgb}{0.09019607843137255, 0.7450980392156863, 0.8117647058823529}

\definecolor{brewer1}{HTML}{e41a1c}
\definecolor{brewer2}{HTML}{377eb8}
\definecolor{brewer3}{HTML}{4daf4a}
\definecolor{brewer4}{HTML}{984ea3}
\definecolor{brewer5}{HTML}{ff7f00}
\definecolor{brewer6}{HTML}{ffff33}
\definecolor{brewer7}{HTML}{a65628}
\definecolor{brewer8}{HTML}{f781bf}
\definecolor{brewer9}{HTML}{999999}
\usepackage{xurl}

\begin{document}
\title{High-precision Quantum Phase Estimation on a Trapped-ion Quantum Computer}
\author{Andrew Tranter}
\email{andrew.tranter@quantinuum.com}
\affiliation{Quantinuum Ltd., Terrington House, 13-15 Hills Road, Cambridge CB2 1NL, United Kingdom}
\author{Duncan Gowland}
\affiliation{Quantinuum Ltd., Terrington House, 13-15 Hills Road, Cambridge CB2 1NL, United Kingdom}
\author{Kentaro Yamamoto}
\affiliation{Quantinuum K.K., Otemachi Financial City Grand Cube 3F, 1-9-2 Otemachi, Chiyoda-ku, Tokyo, Japan}
\author{Michelle Sze}
\affiliation{Quantinuum Ltd., Terrington House, 13-15 Hills Road, Cambridge CB2 1NL, United Kingdom}
\author{David Mu\~noz Ramo}
\affiliation{Quantinuum Ltd., Terrington House, 13-15 Hills Road, Cambridge CB2 1NL, United Kingdom}
\date{\today}
\begin{abstract}
    Emergent quantum computing technologies are widely expected to provide novel approaches in the simulation of quantum chemistry.  Despite rapid improvements in the scale and fidelity of quantum computers, high resource requirements make the execution of quantum chemistry experiments challenging. Typical experiments are limited in the number of qubits used, incur a substantial shot cost, or require complex architecture-specific optimization and error mitigation techniques.

    In this paper, we propose a conceptually simple benchmarking approach involving the use of multi-ancilla quantum phase estimation.  Our approach is restricted to very small chemical systems, and does not scale favorably beyond molecular systems that can be described with $2$ qubits; however, this restriction allows us to generate circuits that scale quadratically in gate count with the number of qubits in the readout register.  This enables the execution of quantum chemistry circuits that act on many qubits, while producing meaningful results with limited shot counts.  We use this technique (with $200$ shots per experiment) to calculate the ground state energy of molecular hydrogen to $50$ bits of precision ($8.9 \times  10^{-16}$ hartree) on a $56$-qubit trapped-ion quantum computer, negating Trotter error.  Including Trotter error, we obtain between $32$ and $36$ bits of precision ($1.5 \times 10^{-10}$ and $6.0 \times 10^{-11}$ hartree respectively), vastly exceeding chemical accuracy ($1.6 \times 10^{-3}$ hartree) against Full Configuration Interaction.  We consider application of the approach to deeper circuits, and discuss potential as a benchmark task for near-term quantum devices.
    \end{abstract}

\maketitle
\newpage
\section{Introduction}

The prospect of highly-accurate quantum chemical simulations at reduced cost is often cited as a promising application of emergent quantum computing devices~\cite{aspuru-guzikSimulatedQuantumComputation2005, mcardleQuantumComputationalChemistry2020a, olsonQuantumInformationComputation2017, QuantumChemistryAge, bauerQuantumAlgorithmsQuantum2020}.  In recent years, the development of such devices has been rapid.  Superconducting devices exceeding $100$ qubits are available for commercial use~\cite{acharyaQuantumErrorCorrection2025, kimEvidenceUtilityQuantum2023}, along with trapped-ion devices for up to $56$ qubits~\cite{chenBenchmarkingTrappedionQuantum2024, mosesRaceTrackTrappedIonQuantum2023}.  While progress has been made in demonstrating quantum error correction techniques on these devices~\cite{acharyaQuantumErrorCorrection2025, eganFaulttolerantControlErrorcorrected2021, paetznickDemonstrationLogicalQubits2024, puttermanHardwareefficientQuantumError2025, yamamotoQuantumErrorCorrectedComputation2025}, application-scale error-corrected experimental work remains firmly out of grasp.

A great deal of methods have been developed in order to perform quantum chemistry simulations on near-term devices~\cite{aspuru-guzikSimulatedQuantumComputation2005, mcardleQuantumComputationalChemistry2020a, olsonQuantumInformationComputation2017, QuantumChemistryAge, bauerQuantumAlgorithmsQuantum2020, kanno2023quantumselectedconfigurationinteractionclassical} -- most significantly, the Variational Quantum Eigensolver (VQE)~\cite{cerezoVariationalQuantumAlgorithms2021, peruzzoVariationalEigenvalueSolver2014a,mccleanTheoryVariationalHybrid2016}. These approaches generally require much shorter circuits than methods designed for error-corrected devices, such as Quantum Phase Estimation (QPE)\cite{kitaevQuantumMeasurementsAbelian1995,abramsSimulationManyBodyFermi1997,nielsenQuantumComputationQuantum2010}. This allows for the simulation of small systems on near-term devices.  Generally, these techniques use a single register of qubits to represent the state of the electrons in the molecule of interest.  It would be desirable to use these methods to assess and benchmark near-term devices with circuits that act upon all of the available qubits; however, the only way to increase the number of qubits used is to increase the complexity of the physical system of interest.  The drawbacks of this are well-known~\cite{gonthierMeasurementsRoadblockNearterm2022}. The depths of the required circuits rapidly outscale the fidelities available on near-term devices.  Similarly, shot counts increase dramatically with the complexity of the system, leading to severe practical limitations in accordance with the restricted availability and speed of near-term devices.  Although in the long term quantum error correction provides a route for addressing both of these problems, the use of commonly-investigated encodings push qubit count to a challenging level, while also still increasing the physical gate count significantly.  In concert, these difficulties lead to the question: in a chemical context, how can we design a benchmark that uses all of the qubits?

Methods to overcome these limitations have been reported, with impressive results~\cite{bauerQuantumAlgorithmsQuantum2020,kimEvidenceUtilityQuantum2023a,googleaiquantumandcollaboratorsHartreeFockSuperconductingQubit2020, robledomoreno2024chemistryexactsolutionsquantumcentric}.  These allow for the simulation of larger molecular systems; however, they are still extremely heavy in required resources, due to sampling expenses.  They also generally require extensive use of error mitigation techniques~\cite{caiQuantumErrorMitigation2023, temmeErrorMitigationShortDepth2017}, which may incur an additional expense (typically shot count).  The performance of such techniques is often dependent on the noise profile of the device, and thus the comparative efficacy of these techniques varies with regards to device architecture.  This additional complexity can impede benchmarking efforts.

In this paper, we propose a conceptually simple alternative.  Rather than attempting to increase qubit use by increasing the complexity of the studied molecular system, we instead restrict ourselves to molecular systems that can be described with only one or two qubits.  While this initially appears counterproductive, it enables the use of fixed-depth decompositions for low-width unitaries.  This allows the meaningful utilization of additional qubits when combined with multi-ancilla quantum phase estimation (QPE).  QPE uses a multi-qubit readout register to obtain an estimate of the desired phase, with resolution dependent on the size of the readout register.  By increasing the size of the readout register, an increasingly precise estimate of the desired energy may be obtained.

Our approach is limited to very small chemical systems, and does not present a scalable avenue to the computation of complex molecular ground-state energies; however, for the purposes of a near-term benchmark, it yields a favourable scaling with regards to the number of qubits used.  This allows us to execute physically-motivated circuits which utilize all of the available qubits with extremely limited shot count, and thus overall cost.

In this paper, we first review multi-ancilla quantum phase estimation and describe our approach to circuit compilation.  We report application of these techniques to the calculation of the ground-state energy of the hydrogen molecule in a minimal basis using a $56$-qubit trapped-ion quantum computer.  We then consider the extension of this technique to a two-qubit Hamiltonian, demonstrating the impact of increasing circuit depth.  Finally, we illustrate the potential of this approach as a hardware benchmark, by reproducing the results on an emulator using a variety of noise models.

\section{High-precision phase estimates}
\label{sec:h2}

Quantum Phase Estimation (QPE) is an algorithm for determining phases accrued by evolution under a given Hamiltonian, and thus the eigenvalue corresponding to a particular eigenvector of this Hamiltonian~\cite{abramsSimulationManyBodyFermi1997, kitaevQuantumMeasurementsAbelian1995, nielsenQuantumComputationQuantum2010}.  A register of \emph{system qubits} is prepared in a state approximating the eigenvector of interest.  A circuit representation of the evolution operator of the Hamiltonian is generated, often using \emph{Trotterization} ~\cite{hatanoFindingExponentialProduct2005, suzukiGeneralizedTrottersFormula1976}, where evolution under each term in the Hamiltonian is performed in sequence, with the number of repetitions (the \emph{Trotter number}) improving the accuracy of the representation.  Sequential binary powers of the evolution operator -- controlled upon the state of a register of \emph{readout qubits} -- are applied, followed by an inverse quantum Fourier transform (\emph{iQFT}) stage.  Assuming the prepared state has good overlap with the target eigenstate, measurement of the readout register carries a high probability of obtaining a value corresponding to the binary decomposition of the desired phase.  The number of qubits in the readout register gives the number of bits of precision on the phase estimate, with the least significant bit often discarded to account for discretization error~\cite{nielsenQuantumComputationQuantum2010}.

\begin{figure}
\begin{quantikz}
\lstick{\ket{+}}   &               & \ctrl{4}\gategroup[5, steps=1, style={dashed, rounded corners, fill=dualAnalagous1!20, inner sep=2pt}, background, label style = {label position=below, anchor=north, yshift=-0.2cm}]{$U^{2^0}_{opt}$}     &                     &                 &       & \gate[4]{iQFT} & \meter{} & \\
\lstick{\ket{+}}   &               &              & \ctrl{3}\gategroup[4, steps=1, style={dashed, rounded corners, fill=dualAnalagous1!20, inner sep=2pt}, background, label style = {label position=above, anchor=north, yshift=0.45cm}]{$(U^{2^0}_{opt})^2$} \gategroup[4, steps=1, style={opacity=0}, label style = {label position=below, anchor=north, yshift=-0.1cm}]{$U^{2^1}_{opt}$}           &                 &          &  & \meter{} & \\
\lstick{\ket{+}}   &               &              &                     & \ctrl{2} \gategroup[3, steps=1, style={dashed, rounded corners, fill=dualAnalagous1!20, inner sep=2pt}, background, label style = {label position=above, anchor=north, yshift=0.45cm}]{$(U^{2^1}_{opt})^2$} \gategroup[3, steps=1, style={opacity=0}, label style = {label position=below, anchor=north, yshift=-0.1cm}]{$U^{2^2}_{opt}$}       &             &  & \meter{} & \\
\lstick{\ket{+}}   &               &              &                     &                 & \ctrl{1} \gategroup[2, steps=1, style={dashed, rounded corners, fill=dualAnalagous1!20, inner sep=2pt}, background, label style = {label position=above, anchor=north, yshift=0.45cm}]{$(U^{2^2}_{opt})^2$} \gategroup[2, steps=1, style={opacity=0}, label style = {label position=below, anchor=north, yshift=-0.1cm}]{$U^{2^3}_{opt}$}         &  & \meter{} & \\
\lstick{\ket{\psi}}&  \qwbundle{N} & \gate{U^{2^0}}   & \gate{U^{2^1}}  &  \gate{U^{2^2}} & \gate{U^{2^3}}               &          &          &
\end{quantikz}
    \caption{A circuit for performing multi-ancilla quantum phase estimation using four readout qubits. Our compilation strategy is to compile each block into an optimized form, and determine the following block by composing it with itself, before compiling the second block, and so on.}
\label{fig:qpe}
\end{figure}
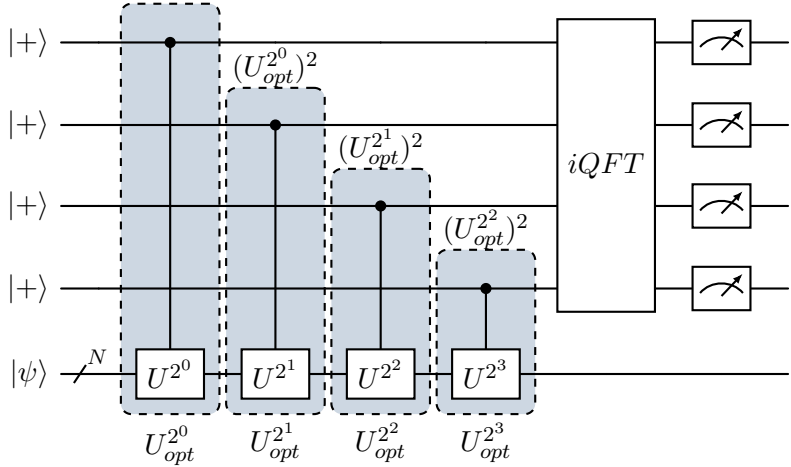

Many novel techniques have been developed~\cite{dobsicekArbitraryAccuracyIterative2007a, obrienQuantumPhaseEstimation2019, svoreFasterPhaseEstimation2013, wiebeEfficientBayesianPhase2016} leveraging other resources to lessen the demanding requirements of this method.  While these methods are likely to be effective for practical implementation at scale, they generally involve the reduction of the readout register to a single qubit, and incur a cost of additional sampling.  Instead, we focus on multi-ancilla phase estimation (as depicted in Fig.~\ref{fig:qpe}), using the size of the readout register to control the quantum computational resources needed for simulation.  An additional benefit of this approach is that it requires no additional sampling to infer phases, and thus requires very few shots to determine the phase of interest.

The computational difficulty of implementing multi-ancilla quantum phase estimation is generally dominated by the application of the $2^{k}$-th power of the unitary of interest, where $k$ is the total number of readout qubits.  This leads to exponential growth in the gate count with respect to $k$, which forms the primary impediment to simulation on near-term devices.  In the completely general case, this problem is difficult to mitigate.  The unitary is determined by the physics encoded in the Hamiltonian of interest, and may not prove amenable to circuit optimization.

The exponential growth in gate count does appear prohibitive; however, the case where the system register is comprised of a limited number of qubits allows for a unique possibility for optimization.  In the case where there are either one or two qubits in the system register, an examination of the circuit in Fig.~\ref{fig:qpe} shows that each controlled $U^{2^{i-1}}$ (where $i$ indexes the $k$ readout qubits) operation acts exclusively on either two or three qubits, depending on whether the system register is of size one or two respectively.  There is $k$ of these two or three qubit subcircuits.  Although the largest subcircuit grows exponentially with $k$, there exist known methods to decompose arbitrary two~\cite{blaauboerAnalyticalDecompositionProtocol2008, vidalUniversalQuantumCircuit2004} and three~\cite{shendeSynthesisQuantumlogicCircuits2006} qubit unitaries to fixed-depth circuits.  If these methods are applied, the pre-iQFT circuit becomes a sequence comprising a linear amount of fixed-depth subcircuits ($U^{2^{i-1}}_{opt}$ in Figure~\ref{fig:qpe}). The overall scaling of gate count then becomes dominated by the iQFT step, which is quadratic with respect to the number of readout qubits.  This is in contrast to typical QPE circuits, wherein the bottleneck is generally the application of the controlled unitary operations.  In the limited case of a one or two qubit system register, it is therefore possible to reduce the typically exponential scaling QPE circuits to a quadratic scaling, which is likely to be much more amenable to implementation on near-term devices.

In practice, generating the initial exponentially large circuit is impractical for currently existing software.  Our approach is therefore:
\begin{enumerate}
    \item Generate an unoptimized subcircuit corresponding to $U^1$.
    \item Compile this subcircuit to an optimized subcircuit, $U^0_{opt}$.
    \item Generate an unoptimized $U^2$ by placing $U^0_{opt}$ twice in sequence.
    \item Compile the $U^2$ circuit to an optimized $U^2_{opt}$.
    \item Repeat until the final $U^{2^{k-1}}_{opt}$ is generated.
\end{enumerate}
From here, the full QPE circuit can be generated, with a final optimization pass applied on the full circuit to ensure a minimal gate count and depth.  As each $U_{opt}$ is of a roughly equal gate count, the largest subcircuit stored at any point is only twice the gate count of an optimized subcircuit.  This approach therefore ensures that no exponential resources are required for circuit compilation.

For generation of the Hamiltonian, the initial QPE circuits and the processing of results, a modified version of the \texttt{InQuanto v3.5} package~\cite{inquanto} was used, along with the \texttt{PySCF v2.5} package~\cite{sunRecentDevelopmentsPySCF2020} and the corresponding \texttt{inquanto-pyscf} extension.  For circuit optimization the \texttt{tket} compiler~\cite{sivarajahT|ketRetargetableCompiler2020} was used -- most notably through the use of the \texttt{FullPeepholeOptimise} pass, which achieves the circuit depth reduction discussed above.

\begin{figure}
\begin{center}
    \begin{tabular}{rl}
 \begin{tikzpicture}[baseline, trim axis left]
  \pgfplotstableread[col sep = comma]{gatecounts.dat}{\gatecountsdat}
  \begin{axis}[
  xlabel={Readout qubits},
  ylabel={Count},
  legend pos = north east,
  ymax = 10000,
  cycle list name=color list,
   ]
  \addplot+ [mark=*,color=dualAnalagous1a]
   table[x={qubits},y={Raw gates}]
    {\gatecountsdat};
   \addplot+ [mark=square*,color=dualAnalagous1b]
  table[x={qubits},y={Raw 2Q gates}]
    {\gatecountsdat};
  \addplot+ [mark=triangle*,color=dualAnalagous1]
  table[x={qubits},y={Raw depth}]
    {\gatecountsdat};
     \addplot+ [mark=*,color=dualAnalagous2a]
  table[x={qubits},y={Optimized gates}]
    {\gatecountsdat};
        \addplot+ [mark=square*,color=dualAnalagous2b]
  table[x={qubits},y={Optimized 2Q gates}]
    {\gatecountsdat};
        \addplot+ [mark=triangle*,color=dualAnalagous2]
  table[x={qubits},y={Optimized depth}]
    {\gatecountsdat};
      \legend{Raw gates, Raw 2Q gates, Raw depth, Optimized gates, Optimized 2Q gates, Optimized depth}
  \end{axis}
 \end{tikzpicture}
        &
         \begin{tikzpicture}[baseline, trim axis right]
               \pgfplotstableread[col sep = comma]{energetics.dat}{\energeticsdat}
  \begin{axis}[
  xlabel={Readout qubits},
  ylabel={Error ($E_h$)},
  legend pos = north east,
  log basis y = {2},
  ymode = log,
  cycle list name=color list
   ]
      \addplot+ [mark=*, color=dualAnalagous1]
  table[x={qubits},y={energy},color=]
    {\energeticsdat};
      \addplot[domain=1:54,
      samples = 100,
      color=dualAnalagous2,
      line width=1pt,
      ]
      {2^(-1 * (x - 2 ))};
    \legend{Experimental, $2^{-(k-2)}$}
      \end{axis}
             \end{tikzpicture}
        \\
\end{tabular}
\end{center}
    \caption{Resources and results of high-width quantum phase estimation for \ce{H2} (STO-3G basis) at equilibrium bond length.  Left: circuit resources by readout qubits used.  Optimization reduces the exponential scaling of the raw circuits to a quadratic scaling, making the circuits amenable to implementation on current devices.  Right: error in the calculated ground-state energy obtained by hardware QPE experiments (blue dots), against a reference value excluding Trotter error ($E_{Trotter}$).  The observed error is equal to or better than the theoretical bound (worst-case discretization error, orange line) until beyond 50 qubits.}
    \label{fig:h2_eqm}
    \end{figure}
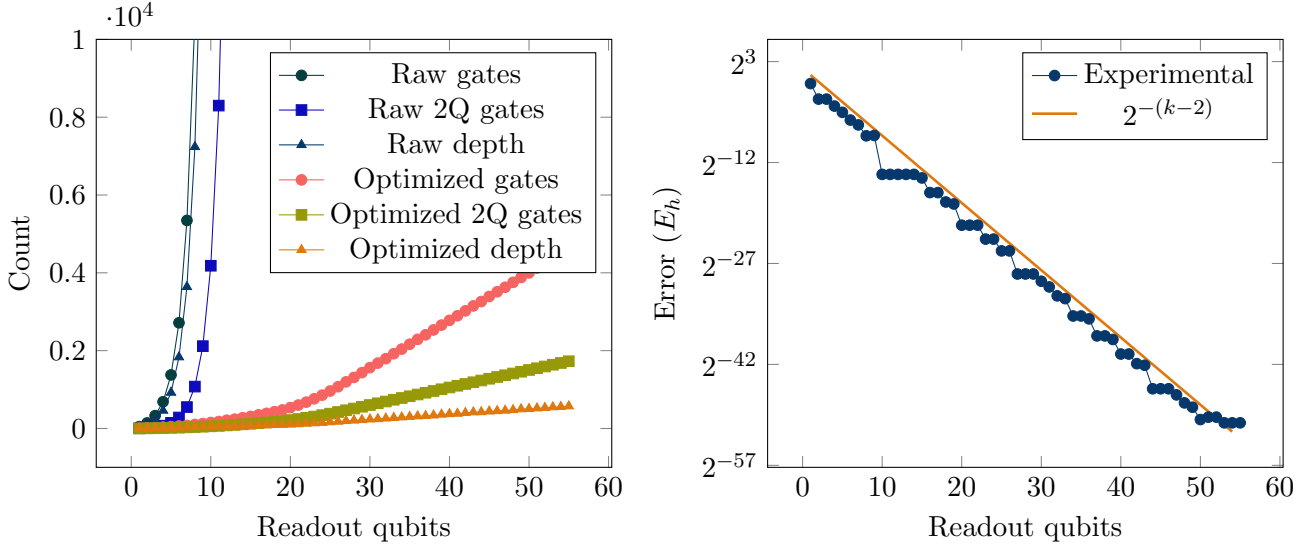

We begin by considering the application of this approach to the molecular hydrogen Hamiltonian in an STO-3G basis.  The fermionic Hamiltonian was mapped to qubits though a standard Jordan-Wigner transformation~\cite{jordanUberPaulischeAquivalenzverbot1928}, and subsequently reduced from 4 to 1 qubit through Z2 tapering~\cite{bravyiTaperingQubitsSimulate2017b, setiaReducingQubitRequirements2020} -- resulting in the Hamiltonian $0.16814576933537334 E_h X + 1.1973374204075313 E_h Z$ after removing identity terms.  QPE circuits for varying numbers of readout qubits were then generated according to the above approach.  The resources of the circuits generated are shown in Fig.~\ref{fig:h2_eqm}.  As expected, we obtain a gate count that scales quadratically with the number of qubits in the readout register.  At 55 qubits in the readout register, the circuit is comprised of $4612$ gates, with a depth of $573$ steps.  These resources are well within the range amenable to currently existing quantum devices.

Following generation and compilation, the circuits were executed on the Quantinuum System Model H2 trapped-ion quantum computer~\cite{mosesRaceTrackTrappedIonQuantum2023}.  For each value of $k$, the most frequent outcome was used to estimate the eigenvalue of interest, with no further inference drawn from the measurement results.  Although this procedure results in a marginally poorer estimate of the phase than would be obtained from further sampling, it allowed an estimate of the phase to be obtained with a minimal amount of shots ($200$ for each experiment).  While the actual accuracy against a reference value is somewhat unpredictable due to discretization effects, we minimize the impact of this by assigning a theoretical bound on the precision on the phase at $k-1$ bits~\cite{nielsenQuantumComputationQuantum2010}.  It is unlikely that sampling limitations or discretization error will result in an error exceeding this bound; instead, such a deviation is likely to be attributable to hardware noise or limitations of the software stack.  As discussed in Appendix~\ref{app:shots}, while the error is generally below this, a tighter bound is likely to be broken in the course of repeated experiments.

When performing QPE, the total evolution time of $U$ may be chosen freely, provided $2 \pi tE$ is within $[0,2 \pi)$.  A shorter evolution time will incur less Trotter error (the error incurred in Trotterization~\cite{childsTheoryTrotterError2021}), but will result in less precision on the eigenvalue calculated.  To obtain a straightforward impact on precision, an evolution time of $0.5 \mathrm{au}$ was used for all experiments.  This resulted in an extra bit of precision loss in the eigenvalue calculation, leading to a theoretical bound on precision on the energy of $k-2$ bits.  The Hartree-Fock state was used as the initial state of the system register, providing sufficient overlap with the true ground state for the equilibrium geometry.  For ease of comparison, no error mitigation, error detection or error correction techniques were used above those implemented at hardware level by the Quantinuum Systems API~\cite{mosesRaceTrackTrappedIonQuantum2023}.  In order to prevent Trotter error  dominating the result, we initially used a single Trotter step to generate an approximation ($U_{Trotter}$) to $U$.  We then compare our energies obtained from QPE against a reference energy ($E_{Trotter}$) obtained through classical diagonalization of $U_{Trotter}$.  $E_{Trotter}$ includes the effect of Trotter error corresponding to the use of one Trotter step; as a result, the calculated error excludes any contribution from Trotter error.  Later in this section, we consider performance against exact diagonalization (Full Configuration Interaction (FCI)), including the use of multiple Trotter steps.

As can be seen on Fig.~\ref{fig:h2_eqm}, the expected $k-2$ bit precision is obtained up to $52$ bits in the phase, corresponding to an error of $8.88 \times 10^{-16}$ hartree, which is substantially beyond chemical precision ($1.6 \times 10^{-3}$ hartree).  The error on the energy exceeds the theoretical bound beyond $50$ readout qubits. We speculate that the failure at $53$ readout qubits is caused by floating-point error in the classical workflow, rather than hardware noise.  Further work is required to test this theory; however, it is supported by an examination of the errors obtained (discussed in Appendix~\ref{app:shots}) and further results discussed in Section~\ref{sec:heh}.

A serendipitous consequence of our approach is that arbitrary Trotter sequences within the controlled evolution subcircuit are also compressed to a constant depth, as each controlled $U$ still acts on only two qubits, regardless of the length of the Trotter sequence.  In principle, this allows for the use of a Trotter approximation with arbitrary numbers of Trotter steps to be performed, for no additional quantum computational cost -- although the increasing classical difficulty of compilation does result in practical limits.  By varying the Trotter approximation used, we may therefore feasibly assess the impact of Trotter error.

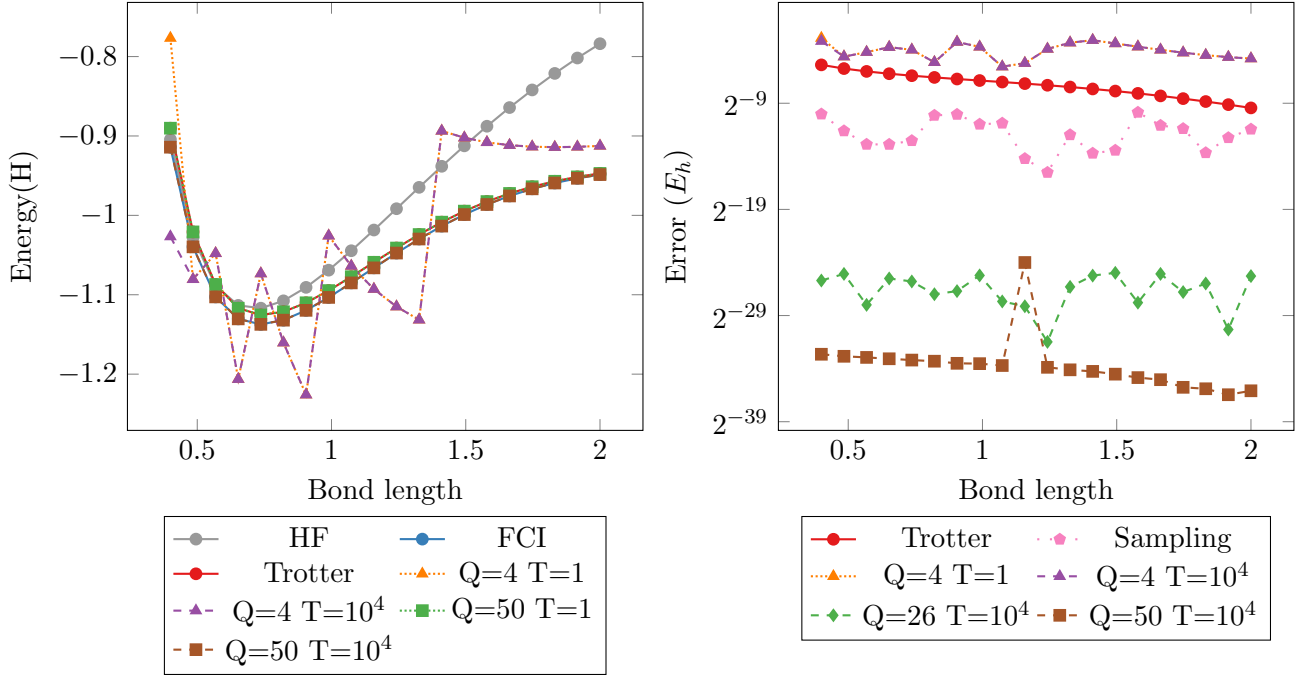
\begin{figure}
\begin{center}
    \begin{tabular}{rl}
 \begin{tikzpicture}[baseline, trim axis left]
  \pgfplotstableread[col sep = comma]{dissociation.dat}{\dissociationdat}
  \begin{axis}[
  xlabel={Bond length},
  ylabel={Energy(H)},
  legend style = {at={(0.5,-0.2)}, anchor=north, legend columns=2},
      cycle list name=exotic,
   ]
  \addplot[mark=*,solid,color=brewer9,line width=0.8pt,mark options={solid}]
   table[x={bondlength},y={hf}]
    {\dissociationdat};
   \addplot[mark=*,solid,color=brewer2,line width=0.8pt,mark options={solid}]
  table[x={bondlength},y={fci}]
    {\dissociationdat};
  \addplot[mark=*,solid,color=brewer1,line width=0.8pt,mark options={solid}]
  table[x={bondlength},y={trotter}]
    {\dissociationdat};
     \addplot[mark=triangle*,densely dotted,color=brewer5,line width=0.8pt,mark options={solid}]
  table[x={bondlength},y={4_1}]
    {\dissociationdat};
        \addplot[mark=triangle*,dashed,color=brewer4,line width=0.8pt,mark options={solid}]
  table[x={bondlength},y={4_10k}]
    {\dissociationdat};
        \addplot[mark=square*,densely dotted,color=brewer3,line width=0.8pt,mark options={solid}]
  table[x={bondlength},y={50_1}]
    {\dissociationdat};
          \addplot[mark=square*,dashed,color=brewer7,line width=0.8pt,mark options={solid}]
  table[x={bondlength},y={50_10k}]
    {\dissociationdat};
      \legend{HF, FCI, Trotter, Q=4 T=1 , Q=4 T=$10^4$, Q=50 T=1, Q=50 T=$10^4$}
  \end{axis}
 \end{tikzpicture}
        &
         \begin{tikzpicture}[baseline, trim axis right]
               \pgfplotstableread[col sep = comma]{dissociation_errors.dat}{\diserrorsdat}
  \begin{axis}[
  xlabel={Bond length},
  ylabel={Error ($E_h$) },
  legend style = {at={(0.5,-0.2)}, anchor=north, legend columns=2},
  log basis y = {2},
  ymode = log,
    cycle list name=exotic
   ]
        \addplot[mark=*,solid,color=brewer1,line width=0.8pt,mark options={solid}]
   table[x={bondlength},y={trotter}]
    {\diserrorsdat};
        \addplot[mark=pentagon*,loosely dotted,color=brewer8,line width=0.8pt,mark options={solid}]
   table[x={bondlength},y={vqe}]
    {\diserrorsdat};
        \addplot[mark=triangle*,densely dotted,color=brewer5,line width=0.8pt,mark options={solid}]
   table[x={bondlength},y={4_1}]
    {\diserrorsdat};
        \addplot[mark=triangle*,dashed,color=brewer4,line width=0.8pt,mark options={solid}]
   table[x={bondlength},y={4_10k}]
      {\diserrorsdat};
    \addplot[mark=diamond*,dashed,color=brewer3,line width=0.8pt,mark options={solid}]
   table[x={bondlength},y={26_10k}]
      {\diserrorsdat};
          \addplot[mark=square*,dashed,color=brewer7,line width=0.8pt,mark options={solid}]
   table[x={bondlength},y={50_10k}]
      {\diserrorsdat};
      \legend{Trotter, Sampling, Q=4 T=1, Q=4 T=$10^4$, Q=26 T=$10^4$, Q=50 T=$10^4$}

      \end{axis}
             \end{tikzpicture}
        \\
\end{tabular}
\end{center}
    \caption{\ce{H2} dissociation results obtained with various levels of readout qubits (Q) and Trotter steps (T).  The sequence labelled ``Trotter" indicates energies obtained from classical diagonalization of the one-Trotter-step evolution operator.  Left: dissociation curves.  The 50 readout qubit examples are indistinguishable from exact curves, and the use of 4 readout qubits does not allow for discrimination between numbers of Trotter steps.  Right: errors obtained as a function of bond length, against FCI.  Errors decrease as more qubits are added to the readout register, and for 50 qubits, the trend suggests that Trotter error dominates.  The sequence labelled ``Sampling" corresponds to an emulated sample-based expectation value measurement experiment, as discussed in the main text.}
    \label{fig:dissocation}
    \end{figure}

Trotter error generally varies with the molecular geometry, and as such it is instructive to consider the full dissociation curve of the molecule.  In the left-hand side of Fig.~\ref{fig:dissocation}, we show dissociation curves for \ce{H2} obtained using our QPE approach on the quantum device, for $4$ and $50$ readout qubits, using $1$ and $10000$ Trotter steps.  At stretched bond lengths, the Hartree-Fock state is a poor approximant to the true ground state; as such, a UCCSD ansatz was used for initial state preparation in these experiments, with optimal coefficients determined by a noiseless emulated VQE calculation~\footnote{For \ce{H2} in an STO-3G basis, an optimized UCCSD ansatz as used here is the exact eigenstate.}.  Results are contrasted against curves generated using Hartree-Fock, Full Configuration Interaction (FCI, providing an ``exact'' reference value), and direct diagonalization of the Trotterized unitary using one Trotter step (i.e. the FCI energy including Trotter error).  Clearly, the $4$ readout qubit experiments do not obtain sufficient accuracy as to effectively represent the dissociation curve.  As the results here are bound by the lack of readout qubits, there is little to distinguish the $1$ and $10000$ Trotter step results, apart from at the shortest bond length, where the Trotter error is at its most substantial.  Using $50$ readout qubits, the opposite result is observed. At the presented scale, the curves for $1$ Trotter step and $10000$ Trotter steps overlap indistinguishably from their respective reference curves (FCI including and excluding one-step Trotter error respectively).  Unlike the $4$ readout qubit case, the impact of Trotter error is readily apparent, with the $10000$ Trotter step circuits producing a markedly more accurate (to FCI) result.

The right-hand side of Fig.~\ref{fig:dissocation} shows the comparative impact of Trotter error and readout imprecision.  We show here the use of $4$, $26$ and $50$ readout qubits, using $10000$ Trotter steps, with error determined against the reference FCI values.  As we increase the number of readout qubits, the error is reduced, in accordance with the higher QPE precision.

With increasing bond length, the Trotter error (red) decreases monotonically.  Conversely, for the $4$ and $26$ qubit QPE experiments, we obtain an error that fluctuates independent of bond length.  This is the expected behaviour from QPE, as discretization error does not have a simple relationship with the physical parameters of the system.  With $50$ readout qubits we obtain a different trend: the error monotonically decreases as bond length increases, the same trend as observed for Trotter error (barring one outlier point attributable to hardware noise, discussed in Appendix~\ref{app:shots}).  The errors obtained are also substantially above the $2^{-48}$ precision obtained from a $50$ readout qubit simulation. It may therefore be inferred that for the $50$ qubit experiments, the error obtained is attributable to Trotter error -- i.e. $10000$ Trotter steps are insufficient to obtain an energy precise to the level expected from this number of readout qubits.  Nonetheless, the obtained accuracy against FCI is substantial: excluding the outlier point mentioned above, we obtain between $32$ and $36$ bits of precision, corresponding to $1.5 \times 10^{-10}$ and $6.0 \times 10^{-11}$ hartree respectively.

To provide further context, a series of noiselessly emulated direct sampling-based expectation value calculations were also performed, in the vein of a typical VQE experiment.  $300000$ shots were used, for each of the $2$ circuits necessary to determine expectation values of the $X$ and $Z$ terms in the Hamiltonian.  As above, state preparation here consisted of a UCCSD ansatz with a precomputed optimal coefficient.  For expediency, these experiments were performed noiselessly on a standard laptop using the \texttt{qiskit-aer} simulator~\cite{javadi-abhariQuantumComputingQiskit2024}, and the corresponding \texttt{pytket-qiskit} \texttt{tket} extension~\cite{CQCLPytketqiskit2025} -- as each circuit was comprised of one single qubit rotation and a measurement, experimental noise was considered to not be a significant factor.  The number of shots was chosen to be very roughly commensurate with the computational cost of the $50$ readout qubit QPE experiments described above, measured in HQC computational units~\cite{QuantinuumSystemsWorkflow}.  Due to unfavourable scaling of precision with shot count, the emulated sampling-based approach incurs greater error than the $50$ readout qubit QPE hardware experiments, by a factor of approximately $2^{20}$.  This comparative accuracy is suggestive of the potential of QPE as a near-term benchmark in the regime wherein available shots are highly limited.

\section{Impact of system register size}
\label{sec:heh}

In Section~\ref{sec:h2}, we presented results focusing on the hydrogen molecule in STO-3G, an example requiring a single system qubit.  This was in order to constrain circuit depth, so as to maximize the chances of success for high-precision experiments involving a large readout register.  The low circuit depth did indeed result in a minimal impact of hardware noise upon the final phase estimate, even when using a near-maximally sized readout register.  While encouraging, this minimal impact of hardware noise impedes analysis of the approach as a hardware benchmark.  In order to better characterise behaviour under noise, it is instructive to consider deeper circuits.

As noted above, our compilation strategy is also effective for system registers comprised of two qubits, as the corresponding circuits will be comprised of three-qubit subcircuits, which are also reducible to a fixed depth -- albeit a substantially higher depth than the one-system-qubit case.  To examine performance within this higher-depth context, we repeated our experiments on the \ce{HeH+}  molecule in STO-3G.  After the application of Z2 tapering~\cite{bravyiTaperingQubitsSimulate2017b, setiaReducingQubitRequirements2020}, this Hamiltonian acts upon a register of two qubits, thus providing the desired circuits.  These experiments were performed with up to $32$ qubits in the readout register.  A Hartree-Fock initial state and one Trotter step were used, with energies again compared against direct diagonalization of the one-Trotter-step unitary.

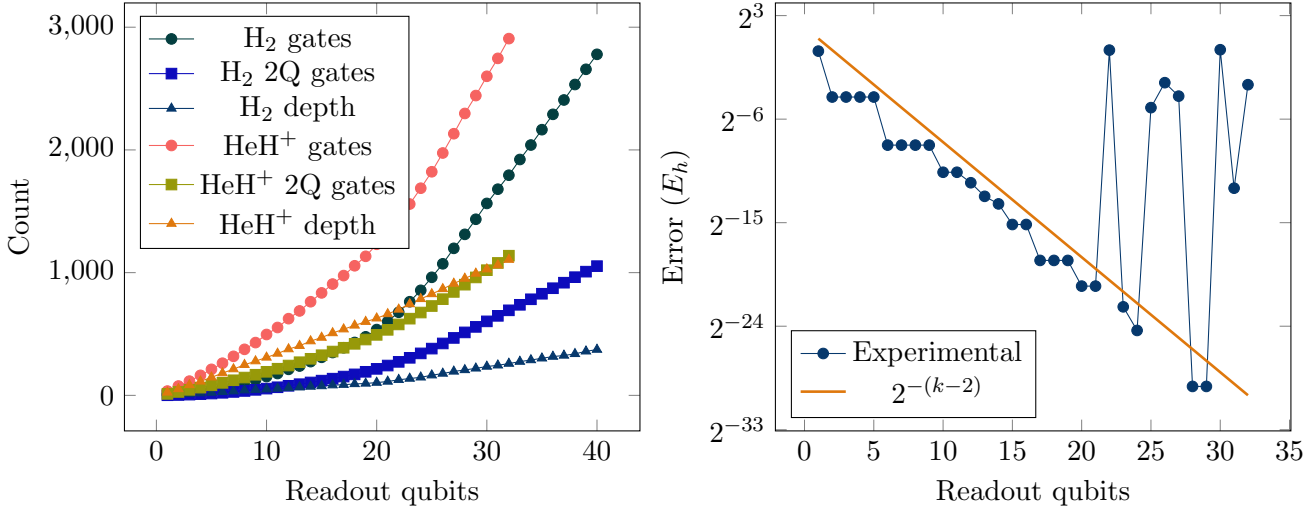
\begin{figure}
\begin{center}
    \begin{tabular}{rl}
 \begin{tikzpicture}[baseline, trim axis left]
     \pgfplotstableread[col sep = comma]{gatecounts.dat}{\gatecountshydat}
  \pgfplotstableread[col sep = comma]{heh_gatecounts.dat}{\gatecountsdat}
  \begin{axis}[
  xlabel={Readout qubits},
  ylabel={Count},
  restrict x to domain=1:40,
  legend pos = north west,
    cycle list name=color list,
   ]
  \addplot+ [mark=*, color=dualAnalagous1a]
   table[x={qubits},y={Optimized gates}]
    {\gatecountshydat};
   \addplot+ [mark=square*,color=dualAnalagous1b]
  table[x={qubits},y={Optimized 2Q gates}]
    {\gatecountshydat};
  \addplot+ [mark=triangle*,color=dualAnalagous1]
  table[x={qubits},y={Optimized depth}]
    {\gatecountshydat};
     \addplot+ [mark=*,color=dualAnalagous2a]
  table[x={qubits},y={Optimized gates}]
    {\gatecountsdat};
        \addplot+ [mark=square*, color=dualAnalagous2b]
  table[x={qubits},y={Optimized 2Q gates}]
    {\gatecountsdat};
        \addplot+ [mark=triangle*, color=dualAnalagous2]
  table[x={qubits},y={Optimized depth}]
    {\gatecountsdat};
      \legend{\ce{H2} gates, \ce{H2} 2Q gates, \ce{H2} depth, \ce{HeH+} gates, \ce{HeH+} 2Q gates, \ce{HeH+} depth}
  \end{axis}
 \end{tikzpicture}
        &
         \begin{tikzpicture}[baseline, trim axis right]
               \pgfplotstableread[col sep = comma]{heh_energetics.dat}{\energeticsdat}
  \begin{axis}[
  xlabel={Readout qubits},
  ylabel={Error ($E_h$)},
  legend pos = south west,
  log basis y = {2},
  ymode = log,
      cycle list name=color list,
   ]
      \addplot+[mark=*,color=dualAnalagous1]
  table[x={qubits},y={energy}]
    {\energeticsdat};
      \addplot[domain=1:32,
      samples = 100,
      color=dualAnalagous2,
      line width=1pt]
      {2^(-1 * (x - 2 ))};
    \legend{Experimental, $2^{-(k-2)}$}
      \end{axis}
             \end{tikzpicture}
        \\
\end{tabular}
\end{center}
    \caption{Resources and results of high-width quantum phase estimation for \ce{HeH+}.  Left: optimized circuit resources by readout qubits used, with the resources for \ce{H2} included for comparison.  The depth increases faster for \ce{HeH+} than for \ce{H2}, reaching approximately $1000$ steps by $30$ readout qubits.  Right: error in the calculated ground-state energy obtained by hardware QPE experiments (blue dots) on \ce{HeH+}, against a reference value excluding Trotter error ($E_{Trotter}$).  The results are dominated by hardware noise beyond $20$ qubits in the readout register.}
    \label{fig:heh}
    \end{figure}

The resources required for these simulations are shown on the left-hand side of Figure~\ref{fig:heh}, with the obtained error shown on the right-hand side.  As to be expected, there is substantially faster growth in resources when compared to the experiments discussed in Section~\ref{sec:h2}.  The circuits suffer significantly from a more limited degree of gate parallelizability, with the depth being comparatively much closer to the total gate count than for the experiments discussed in Section~\ref{sec:h2}.

As a consequence of this increased circuit depth, the results for these experiments begin to fail at a much earlier point, as shown on the right-hand side of Figure~\ref{fig:heh}. By $22$ qubits in the readout register, the results are unreliable, with varying degrees of error.  The failure of these experiments is attributable to hardware noise incurred from the deeper circuits.  The errors may be contrasted to those observed for the hydrogen experiments (depicted in Figure~\ref{fig:h2_eqm}).  In the \ce{HeH+} results, the experiments where error exceeds the theoretical bound ($2^{-(k-2)}$), exceed it by many orders of magnitude.  In Figure~\ref{fig:h2_eqm}, the opposite may be observed: the errors which exceed the theoretical error bound are still close to the bound.  This points to different modes of failure between the experiments, lending credence to our earlier suggestion that failure of the largest experiments of Section~\ref{sec:h2} is due to limitations outside of hardware noise.

In this paper, we have restricted ourselves to the consideration of one- and two-qubit systems.  This is due to the ability of the \texttt{tket} compiler to easily compress two and three-qubit unitaries through built-in optimization passes.  It should be noted that many decomposition techniques exist for unitaries acting on greater numbers of qubits~\cite{barencoElementaryGatesQuantum1995,krolHighlyEfficientDecomposition2024, rakytaApproachingTheoreticalLimit2022, shendeSynthesisQuantumlogicCircuits2006, vartiainenEfficientDecompositionQuantum2004}.  In general, these scale exponentially with the number of qubits, and thus do not provide a route to scalability in the system register size.  Nevertheless, it is likely that our approach could be extended to mildly larger systems.  This could present a useful direction when developing benchmarks for future, higher-fidelity devices.

\section{Use as a hardware benchmark}
\label{sec:benchmarking}

As the circuit may easily be systematically constructed and the result verified, the approach described in this paper can be effectively utilized as a device benchmark.  In prior sections of this paper we have reported the results of our experiments in terms of the error obtained on the ground-state energy.  This is instructive with regards to the main purpose of the algorithm; however, for use as a hardware benchmark, it is more useful to consider the impact of increasing readout qubit count upon the strength of the signal obtained.

The degree as to which hardware noise effects the observed signal may be assessed by considering two factors. The first is the probability of obtaining the most frequent outcome ($P_{max}$).  This requires no knowledge of the true eigenvalue.  $P_{max}$ will generally be lower than $1$ due to discretization error, but a qualitative trend can be determined.  Furthermore, complete decoherence would lead to approximately uniformly sampled bitstrings with $P_{max}=1/N_{shots}$, where $N_{shots}$ is the shot count. Any significant deviation of $P_{max}$ from $1/N_{shots}$ for large $k$ indicates a degree of successful algorithmic execution, particularly as the number of possible bitstrings grows exponential in $k$.  The second factor is whether the most frequent outcome obtained (with probability $P_{max}$) is the ``correct" bitstring, corresponding to the true eigenvalue (i.e. the phase estimation procedure successfully estimated the correct phase).

\begin{figure}
\begin{center}
    \begin{tabular}{rl}
 \begin{tikzpicture}[baseline, trim axis left]
  \pgfplotstableread[col sep = comma]{maxprob_h2.dat}{\maxprobdat}
  \begin{axis}[
  ybar=0.5pt,
  xlabel={Readout qubits},
  ylabel={$P_{max}$},
  legend pos = north east,
  bar width = 0.01cm,
  enlarge x limits = 0.05,
   ]
  \addplot+[color=triad3]
   table[x={qubits},y={statevector}]
    {\maxprobdat};
   \addplot+[color=triad2]
  table[x={qubits},y={noiseless}]
    {\maxprobdat};
  \addplot+[color=triad1]
  table[x={qubits},y={h2}]
    {\maxprobdat};
      \legend{Statevector,Noiseless,Hardware}
  \end{axis}
 \end{tikzpicture}
        &
 \begin{tikzpicture}[baseline, trim axis right]
  \pgfplotstableread[col sep = comma]{maxprob_heh.dat}{\maxprobhehdat}
  \begin{axis}[
  ybar=0.5pt,
  xlabel={Readout qubits},
  ylabel={$P_{max}$},
  legend pos = north east,
  bar width = 0.01cm,
  enlarge x limits = 0.05,
   ]
  \addplot+[color=triad3]
   table[x={qubits},y={statevector}]
    {\maxprobhehdat};
   \addplot+[color=triad2]
  table[x={qubits},y={noiseless}]
    {\maxprobhehdat};
  \addplot+[color=triad1]
  table[x={qubits},y={h2}]
    {\maxprobhehdat};
      \legend{Statevector, Noiseless, Hardware}
  \end{axis}
 \end{tikzpicture}
        \\
\end{tabular}
\end{center}
    \caption{Decay of $P_{max}$ increasing number of readout qubits.  $P_{max}$ is the probability of obtaining the most likely outcome (i.e. the success probability).  Left: Simulation of \ce{H2}.  $P_{max}$ decays to a low level in accordance with increasingly large circuits.  Right: Simulation of \ce{HeH+}.  The drop-off in signal occurs faster due to increased hardware noise, leading to the erroneous results.}
    \label{fig:pmax}
    \end{figure}
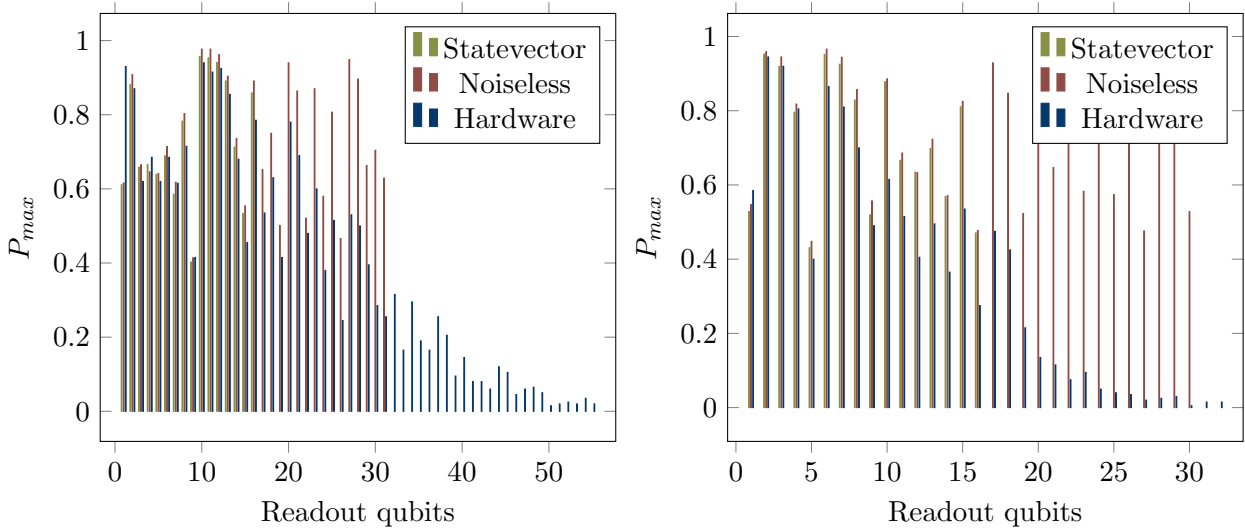

A comparison to noiseless simulation proves instructive. In Figure~\ref{fig:pmax}, we plot the probability of reading out the most frequent outcome, $P_{max}$, against readout qubit count.  Whilst the results for \ce{H2} in Section~\ref{sec:h2} show accuracy for up to $50$ readout qubits, we do observe a substantial drop in the signal as the readout qubit count increases. By $50$ readout qubits, the signal is clearly very weak, with only a few shots (out of $200$) obtaining the ``correct'' result.  Due to increased depth and gate count, the \ce{HeH+} experiments exacerbate this behaviour, with the signal becoming very low by $30$ readout qubits.  This rapid decay helps explain the erroneous phase predictions at high readout qubit counts, shown in Fig.~\ref{fig:heh}: the probability to read the correct bitstring has reduced enough such that other, incorrect results are obtained with equal or larger probability.

In order to serve as a device benchmark, our method should display noise sensitivities. We demonstrate this through emulation of the H2 trapped-ion device with scalings applied to the error model parameters (see~\cite{SystemModelH2} for error model details). We note that our two example systems appear to span very well the range of quasi-noiseless to almost entirely noisy results for current device fidelities, with qubit counts ranging from 2 to 30 and depths of between 50 and 1000 2-qubit gates. These ranges are sufficiently small to be accurately modeled with high fidelity device emulation.

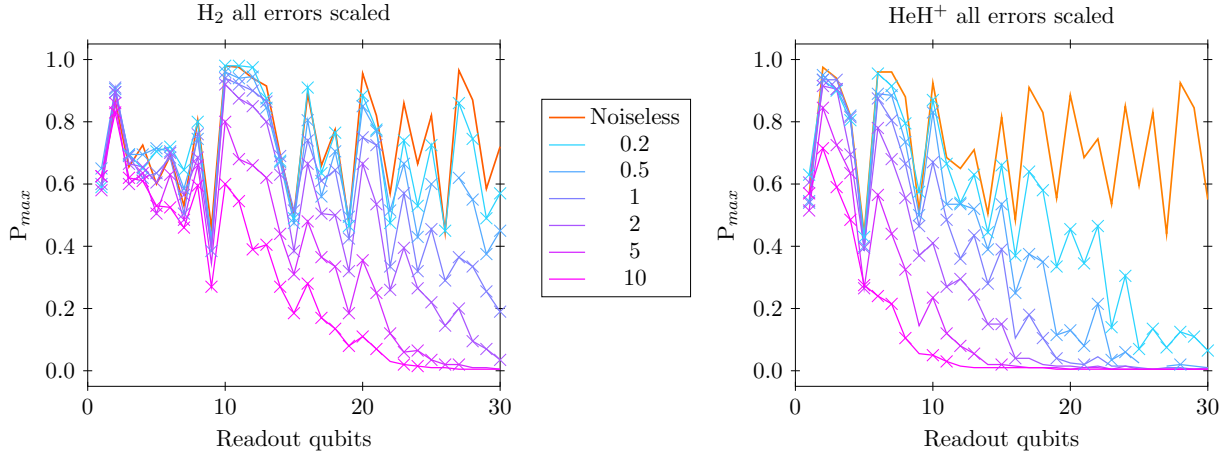
\begin{figure}[h]
    \centering
\begin{tikzpicture}[scale=0.8]

\definecolor{cornflowerblue85170255}{RGB}{85,170,255}
\definecolor{cyan}{RGB}{0,255,255}
\definecolor{darkgray176}{RGB}{176,176,176}
\definecolor{deepskyblue42213255}{RGB}{42,213,255}
\definecolor{magenta}{RGB}{255,0,255}
\definecolor{magenta21341255}{RGB}{213,41,255}
\definecolor{mediumorchid17085255}{RGB}{170,85,255}
\definecolor{mediumslateblue128127255}{RGB}{128,127,255}

\begin{axis}[
tick align=center,
tick pos=both,
legend style={at={(1.1,0.55)},anchor=west,},
title={H$_2$ all errors scaled},
unbounded coords=jump,
x grid style={darkgray176},
xlabel={Readout qubits},
xmin=0, xmax=30,
xtick style={color=black},
xtick={0,10,20,30},
xticklabels={
  \(\displaystyle {0}\),
  \(\displaystyle {10}\),
  \(\displaystyle {20}\),
  \(\displaystyle {30}\)
},
y grid style={darkgray176},
ylabel={P\(\displaystyle _{max}\)},
ymin=-0.05, ymax=1.05,
ytick style={color=black},
ytick={-0.2,0,0.2,0.4,0.6,0.8,1,1.2},
yticklabels={
  \(\displaystyle {\ensuremath{-}0.2}\),
  \(\displaystyle {0.0}\),
  \(\displaystyle {0.2}\),
  \(\displaystyle {0.4}\),
  \(\displaystyle {0.6}\),
  \(\displaystyle {0.8}\),
  \(\displaystyle {1.0}\),
  \(\displaystyle {1.2}\)
}
]
\addplot [draw=deepskyblue42213255, fill=deepskyblue42213255, mark=x,mark size=4pt, only marks]
table{%
x  y
30 0.57
29 0.49
28 0.745
27 0.86
26 0.45
25 0.725
24 0.53
23 0.74
22 0.475
21 0.775
20 0.885
19 0.465
18 0.765
17 0.635
16 0.91
15 0.49
14 0.675
13 0.875
12 0.975
11 0.98
10 0.98
nan nan
8 0.8
7 0.645
6 0.72
5 0.705
4 0.65
3 0.695
2 0.895
1 0.59
};
\addplot [draw=cornflowerblue85170255, fill=cornflowerblue85170255, mark=x,mark size=4pt, only marks]
table{%
x  y
30 0.45
29 0.375
28 0.55
27 0.62
nan nan
25 0.6
24 0.43
23 0.665
22 0.515
21 0.77
20 0.85
19 0.495
18 0.705
17 0.56
16 0.805
15 0.475
14 0.67
13 0.845
12 0.945
11 0.94
10 0.96
nan nan
8 0.745
7 0.585
6 0.705
5 0.715
4 0.695
3 0.695
2 0.905
1 0.65
};
\addplot [draw=mediumslateblue128127255, fill=mediumslateblue128127255, mark=x,mark size=4pt, only marks]
table{%
x  y
30 0.19
29 0.255
28 0.335
27 0.365
26 0.29
25 0.455
24 0.32
23 0.57
22 0.335
21 0.725
20 0.75
19 0.425
18 0.645
17 0.61
16 0.74
15 0.515
14 0.69
13 0.865
12 0.9
11 0.92
10 0.94
9 0.385
8 0.76
7 0.505
6 0.685
5 0.635
4 0.62
3 0.69
2 0.91
1 0.6
};
\addplot [draw=mediumorchid17085255, fill=mediumorchid17085255, mark=x,mark size=4pt, only marks]
table{%
x  y
30 0.0349999999999999
29 0.0700000000000001
28 0.095
27 0.2
26 0.145
25 0.22
24 0.265
23 0.395
22 0.26
21 0.535
20 0.665
19 0.32
18 0.5
17 0.505
16 0.665
15 0.385
14 0.63
13 0.8
12 0.85
11 0.875
10 0.92
nan nan
8 0.685
7 0.565
6 0.7
5 0.605
4 0.655
3 0.665
2 0.895
1 0.625
};
\addplot [draw=magenta21341255, fill=magenta21341255, mark=x,mark size=4pt, only marks]
table{%
x  y
27 0.02
26 0.02
25 0.0349999999999999
24 0.0649999999999999
23 0.0600000000000001
22 0.12
21 0.25
20 0.355
19 0.185
18 0.335
17 0.365
16 0.48
15 0.31
14 0.44
13 0.62
12 0.66
11 0.68
10 0.8
nan nan
8 0.66
7 0.485
6 0.63
5 0.505
4 0.635
3 0.6
2 0.86
1 0.58
};
\addplot [draw=magenta, fill=magenta, mark=x,mark size=4pt, only marks]
table{%
x  y
24 0.0149999999999999
23 0.02
nan nan
21 0.0700000000000001
20 0.11
19 0.0800000000000001
18 0.135
17 0.17
16 0.28
15 0.185
14 0.27
13 0.405
12 0.39
11 0.545
10 0.6
9 0.27
8 0.595
7 0.46
6 0.525
5 0.53
4 0.61
3 0.62
2 0.83
1 0.625
};
\addplot [thick, neonorange]
table {%
30 0.720000028610229
29 0.585000038146973
28 0.870000004768372
27 0.964999914169312
26 0.444999933242798
25 0.819999933242798
24 0.664999961853027
23 0.860000014305115
22 0.569999933242798
21 0.819999933242798
20 0.954999923706055
19 0.475000023841858
18 0.774999976158142
17 0.660000085830688
16 0.889999985694885
15 0.5
14 0.694999933242798
13 0.914999961853027
12 0.940000057220459
11 0.975000023841858
10 0.980000019073486
9 0.460000038146973
8 0.805000066757202
7 0.529999971389771
6 0.690000057220459
5 0.600000023841858
4 0.725000023841858
3 0.654999971389771
2 0.865000009536743
1 0.600000023841858
};
\addplot [semithick, deepskyblue42213255]
table {%
30 0.569999933242798
29 0.490000009536743
28 0.745000004768372
27 0.860000014305115
26 0.450000047683716
25 0.725000023841858
24 0.529999971389771
23 0.740000009536743
22 0.475000023841858
21 0.774999976158142
20 0.884999990463257
19 0.465000033378601
18 0.764999985694885
17 0.634999990463257
16 0.910000085830688
15 0.490000009536743
14 0.674999952316284
13 0.875
12 0.975000023841858
11 0.980000019073486
10 0.980000019073486
9 0.409999966621399
8 0.799999952316284
7 0.644999980926514
6 0.720000028610229
5 0.704999923706055
4 0.649999976158142
3 0.694999933242798
2 0.894999980926514
1 0.589999914169312
};
\addplot [semithick, cornflowerblue85170255]
table {%
30 0.450000047683716
29 0.375
28 0.549999952316284
27 0.620000004768372
nan nan
25 0.600000023841858
24 0.430000066757202
23 0.664999961853027
22 0.514999985694885
21 0.769999980926514
20 0.850000023841858
19 0.495000004768372
18 0.704999923706055
17 0.559999942779541
16 0.805000066757202
15 0.475000023841858
14 0.670000076293945
13 0.845000028610229
12 0.944999933242798
11 0.940000057220459
10 0.960000038146973
9 0.424999952316284
8 0.745000004768372
7 0.585000038146973
6 0.704999923706055
5 0.714999914169312
4 0.694999933242798
3 0.694999933242798
2 0.904999971389771
1 0.649999976158142
};
\addplot [semithick, mediumslateblue128127255]
table {%
30 0.190000057220459
29 0.254999995231628
28 0.335000038146973
27 0.365000009536743
26 0.289999961853027
25 0.455000042915344
24 0.319999933242798
23 0.569999933242798
22 0.335000038146973
21 0.725000023841858
20 0.75
19 0.424999952316284
18 0.644999980926514
17 0.610000014305115
16 0.740000009536743
15 0.514999985694885
13 0.865000009536743
12 0.899999976158142
10 0.940000057220459
9 0.384999990463257
8 0.759999990463257
7 0.504999995231628
6 0.684999942779541
5 0.634999990463257
4 0.620000004768372
3 0.690000057220459
2 0.910000085830688
1 0.600000023841858
};
\addplot [semithick, mediumorchid17085255]
table {%
30 0.0349999666213989
29 0.0700000524520874
28 0.0950000286102295
27 0.200000047683716
26 0.144999980926514
25 0.220000028610229
24 0.264999985694885
23 0.394999980926514
22 0.259999990463257
21 0.535000085830688
20 0.664999961853027
19 0.319999933242798
18 0.5
17 0.504999995231628
16 0.664999961853027
15 0.384999990463257
14 0.629999995231628
13 0.799999952316284
12 0.850000023841858
11 0.875
10 0.920000076293945
9 0.394999980926514
8 0.684999942779541
7 0.565000057220459
6 0.700000047683716
5 0.605000019073486
4 0.654999971389771
3 0.664999961853027
2 0.894999980926514
1 0.625
};
\addplot [semithick, magenta21341255]
table {%
30 0.00499999523162842
29 0.00999999046325684
28 0.00999999046325684
27 0.0199999809265137
26 0.0199999809265137
25 0.0349999666213989
24 0.065000057220459
23 0.059999942779541
22 0.120000004768372
21 0.25
20 0.355000019073486
19 0.184999942779541
18 0.335000038146973
17 0.365000009536743
16 0.480000019073486
15 0.309999942779541
14 0.440000057220459
13 0.620000004768372
12 0.660000085830688
11 0.680000066757202
10 0.799999952316284
9 0.345000028610229
8 0.660000085830688
7 0.485000014305115
6 0.629999995231628
5 0.504999995231628
4 0.634999990463257
3 0.600000023841858
2 0.860000014305115
1 0.579999923706055
};
\addplot [semithick, magenta]
table {%
30 0.00499999523162842
27 0.00499999523162842
26 0.00999999046325684
25 0.00999999046325684
23 0.0199999809265137
22 0.0299999713897705
20 0.110000014305115
19 0.0800000429153442
18 0.134999990463257
17 0.169999957084656
16 0.279999971389771
15 0.184999942779541
14 0.269999980926514
13 0.404999971389771
12 0.389999985694885
11 0.545000076293945
10 0.600000023841858
9 0.269999980926514
8 0.595000028610229
7 0.460000038146973
6 0.524999976158142
5 0.529999971389771
4 0.610000014305115
3 0.620000004768372
2 0.829999923706055
1 0.625
};
\legend{ , , , , , ,Noiseless,0.2, 0.5, 1, 2,5,10}

\end{axis}

\end{tikzpicture}
\begin{tikzpicture}[scale=0.8]

\definecolor{cornflowerblue85170255}{RGB}{85,170,255}
\definecolor{cyan}{RGB}{0,255,255}
\definecolor{darkgray176}{RGB}{176,176,176}
\definecolor{deepskyblue42213255}{RGB}{42,213,255}
\definecolor{magenta}{RGB}{255,0,255}
\definecolor{magenta21341255}{RGB}{213,41,255}
\definecolor{mediumorchid17085255}{RGB}{170,85,255}
\definecolor{mediumslateblue128127255}{RGB}{128,127,255}

\begin{axis}[
tick align=center,
tick pos=both,
title={HeH$^+$ all errors scaled},
unbounded coords=jump,
x grid style={darkgray176},
xlabel={Readout qubits},
xmin=0, xmax=30,
xtick style={color=black},
xtick={0,10,20,30},
xticklabels={
  \(\displaystyle {0}\),
  \(\displaystyle {10}\),
  \(\displaystyle {20}\),
  \(\displaystyle {30}\)
},
y grid style={darkgray176},
ylabel={P\(\displaystyle _{max}\)},
ymin=-0.05, ymax=1.05,
ytick style={color=black},
ytick={-0.2,0,0.2,0.4,0.6,0.8,1,1.2},
yticklabels={
  \(\displaystyle {\ensuremath{-}0.2}\),
  \(\displaystyle {0.0}\),
  \(\displaystyle {0.2}\),
  \(\displaystyle {0.4}\),
  \(\displaystyle {0.6}\),
  \(\displaystyle {0.8}\),
  \(\displaystyle {1.0}\),
  \(\displaystyle {1.2}\)
}
]
\addplot [draw=deepskyblue42213255, fill=deepskyblue42213255, mark=x,mark size=4pt,mark size=4pt, only marks]
table{%
x  y
30 0.0649999999999999
29 0.11
28 0.125
27 0.075
26 0.135
25 0.0700000000000001
24 0.305
23 0.14
22 0.465
21 0.345
20 0.455
19 0.335
18 0.58
17 0.64
16 0.37
15 0.66
14 0.445
13 0.63
12 0.54
11 0.665
10 0.87
9 0.575
8 0.795
7 0.915
6 0.955
5 0.43
4 0.805
3 0.9
2 0.935
1 0.63
};
\addplot [draw=cornflowerblue85170255, fill=cornflowerblue85170255, mark=x,mark size=4pt,mark size=4pt, only marks]
table{%
x  y
28 0.02
nan nan
24 0.0600000000000001
23 0.0349999999999999
22 0.215
21 0.0800000000000001
20 0.13
19 0.115
18 0.35
17 0.375
16 0.25
15 0.535
14 0.39
13 0.52
12 0.535
11 0.535
10 0.83
9 0.5
8 0.735
7 0.885
6 0.89
5 0.4
4 0.82
3 0.905
2 0.95
1 0.545
};
\addplot [draw=mediumslateblue128127255, fill=mediumslateblue128127255, mark=x,mark size=4pt,mark size=4pt, only marks]
table{%
x  y
21 0.02
nan nan
19 0.04
18 0.105
17 0.18
nan nan
15 0.39
14 0.28
13 0.435
12 0.36
11 0.485
10 0.67
9 0.465
8 0.665
7 0.805
6 0.875
nan nan
4 0.815
3 0.935
2 0.935
1 0.54
};
\addplot [draw=mediumorchid17085255, fill=mediumorchid17085255, mark=x,mark size=4pt,mark size=4pt, only marks]
table{%
x  y
16 0.04
15 0.15
14 0.15
13 0.245
12 0.295
11 0.27
10 0.41
9 0.37
8 0.555
7 0.68
6 0.78
5 0.395
4 0.695
3 0.905
2 0.915
1 0.605
};
\addplot [draw=magenta21341255, fill=magenta21341255, mark=x,mark size=4pt,mark size=4pt, only marks]
table{%
x  y
15 0.02
nan nan
13 0.0549999999999999
12 0.0800000000000001
11 0.12
10 0.235
nan nan
8 0.325
7 0.44
6 0.565
5 0.265
4 0.635
3 0.725
2 0.845
1 0.515
};
\addplot [draw=magenta, fill=magenta, mark=x,mark size=4pt,mark size=4pt, only marks]
table{%
x  y
11 0.03
10 0.05
nan nan
8 0.105
7 0.215
6 0.24
5 0.275
4 0.485
3 0.59
2 0.715
1 0.57
};
\addplot [thick, orange]
table {%
30 0.549999952316284
29 0.845000028610229
28 0.924999952316284
27 0.440000057220459
26 0.829999923706055
25 0.595000028610229
24 0.850000023841858
23 0.535000085830688
22 0.745000004768372
21 0.684999942779541
20 0.884999990463257
19 0.555000066757202
18 0.829999923706055
17 0.910000085830688
16 0.485000014305115
15 0.815000057220459
14 0.504999995231628
13 0.710000038146973
12 0.649999976158142
11 0.684999942779541
10 0.924999952316284
9 0.514999985694885
8 0.879999995231628
7 0.960000038146973
6 0.960000038146973
5 0.440000057220459
4 0.819999933242798
3 0.940000057220459
2 0.975000023841858
1 0.565000057220459
};
\addplot [semithick, deepskyblue42213255]
table {%
30 0.065000057220459
29 0.110000014305115
28 0.125
27 0.0750000476837158
26 0.134999990463257
25 0.0700000524520874
24 0.305000066757202
23 0.139999985694885
22 0.465000033378601
21 0.345000028610229
20 0.455000042915344
19 0.335000038146973
18 0.579999923706055
17 0.639999985694885
16 0.370000004768372
15 0.660000085830688
14 0.444999933242798
13 0.629999995231628
12 0.539999961853027
11 0.664999961853027
10 0.870000004768372
9 0.575000047683716
8 0.795000076293945
7 0.914999961853027
6 0.954999923706055
5 0.430000066757202
4 0.805000066757202
3 0.899999976158142
2 0.934999942779541
1 0.629999995231628
};
\addplot [semithick, cornflowerblue85170255]
table {%
30 0.00999999046325684
28 0.0199999809265137
27 0.0149999856948853
nan nan
25 0.0249999761581421
24 0.059999942779541
23 0.0349999666213989
22 0.215000033378601
21 0.0800000429153442
20 0.129999995231628
19 0.115000009536743
18 0.350000023841858
17 0.375
16 0.25
15 0.535000085830688
14 0.389999985694885
13 0.519999980926514
12 0.535000085830688
11 0.535000085830688
10 0.829999923706055
9 0.5
8 0.735000014305115
7 0.884999990463257
6 0.889999985694885
5 0.399999976158142
4 0.819999933242798
3 0.904999971389771
2 0.950000047683716
1 0.545000076293945
};
\addplot [semithick, mediumslateblue128127255]
table {%
30 0.00999999046325684
29 0.00499999523162842
28 0.00999999046325684
27 0.00999999046325684
26 0.00499999523162842
24 0.0149999856948853
23 0.0149999856948853
22 0.0449999570846558
21 0.0199999809265137
20 0.0249999761581421
19 0.0399999618530273
18 0.105000019073486
17 0.180000066757202
16 0.105000019073486
15 0.389999985694885
14 0.279999971389771
13 0.434999942779541
12 0.360000014305115
11 0.485000014305115
10 0.670000076293945
9 0.465000033378601
8 0.664999961853027
7 0.805000066757202
6 0.875
5 0.384999990463257
4 0.815000057220459
3 0.934999942779541
2 0.934999942779541
1 0.539999961853027
};
\addplot [semithick, mediumorchid17085255]
table {%
30 0.00499999523162842
25 0.00499999523162842
24 0.0149999856948853
23 0.00499999523162842
22 0.0149999856948853
21 0.00999999046325684
20 0.0149999856948853
19 0.0149999856948853
18 0.0199999809265137
17 0.0399999618530273
16 0.0399999618530273
15 0.149999976158142
14 0.149999976158142
13 0.245000004768372
12 0.294999957084656
11 0.269999980926514
10 0.409999966621399
9 0.370000004768372
8 0.555000066757202
7 0.680000066757202
6 0.779999971389771
5 0.394999980926514
4 0.694999933242798
3 0.904999971389771
2 0.914999961853027
1 0.605000019073486
};
\addplot [semithick, magenta21341255]
table {%
30 0.00499999523162842
23 0.00499999523162842
22 0.00999999046325684
21 0.00999999046325684
20 0.00499999523162842
19 0.00999999046325684
17 0.00999999046325684
15 0.0199999809265137
14 0.0199999809265137
13 0.0549999475479126
12 0.0800000429153442
11 0.120000004768372
10 0.235000014305115
9 0.144999980926514
8 0.325000047683716
7 0.440000057220459
6 0.565000057220459
5 0.264999985694885
4 0.634999990463257
3 0.725000023841858
2 0.845000028610229
1 0.514999985694885
};
\addplot [semithick, magenta]
table {%
30 0.00499999523162842
19 0.00499999523162842
18 0.00999999046325684
13 0.00999999046325684
12 0.0149999856948853
11 0.0299999713897705
10 0.0499999523162842
9 0.0549999475479126
8 0.105000019073486
7 0.215000033378601
6 0.240000009536743
5 0.274999976158142
4 0.485000014305115
3 0.589999914169312
2 0.714999914169312
1 0.569999933242798
};

\end{axis}

\end{tikzpicture}
    \caption{Probability to measure the most likely phase for between 1 and 30 readout ancillae as a function of the H2 ion-trap emulator error model scaling ($\times 0$ (noiseless) at top in blue $\rightarrow$ $\times 10$ at bottom in pink) . `$\times$' marks denote results where the highest probability phase of the noisy result matches the highest probability phase of the noiseless result for 200 shots. }
    \label{all_error_scaled}
\end{figure}

In Figure~\ref{all_error_scaled}, we demonstrate the decay of the phase readout probability as a function of scaling all noise model parameters. For example, a value of 10 represents making all model errors ``10 times worse'', by increasing the probability for a 2-qubit gate error by a factor of 10, increasing the rate at which qubit dephasing occurs, and other such factors (measurement error, crosstalk, etc.). Conversely, an error scale below 1 represents a less noisy device than the H2 system, across all error model parameters. For \ce{HeH+}, the emulator results with an error scaling of $1.0$ replicate the hardware results in Figure~\ref{fig:heh}, yielding incorrect phases above 21 readout qubits. This suggests that the emulator is a good facsimile of the hardware device. As expected, increasing errors leads to a decay in the QPE signal, resulting in reading out many bitstrings with low probability. However, decaying the signal sufficiently such that the highest probability phase measured is incorrect for this set of readout qubits, requires significant increase in noise model parameters: the QPE protocol appears quite noise-resilient. As the noise parameters are increased, we observe a smoothening of $P_{max}$, as the discretization error becomes small relative to errors from noise.  In supplemental Figure~\ref{allscaled_p3max}, we present smoother decay of QPE readout by examining the sum of the three highest probability phases. This approach obviates some discretization error and more precisely highlights the effect of device noise.

We further inspected the role of specific errors by scaling the 2-qubit gate error and the `memory error', whilst keeping all other parameters as the device benchmarked values. These results are presented and discussed in the supplemental material. To summarize those results: even for large specific error scalings ($\times 100$) the algorithm succeeds for \ce{H2} for more than 12 readout qubits, and we find greater sensitivity to memory than two-qubit gate error. Similar comments regarding memory error were made in recent error-corrected phase estimation experiments of chemical systems~\cite{yamamotoQuantumErrorCorrectedComputation2025}.

Our experiments using the device emulator with two model systems demonstrate the noise sensitivities in our approach. Whilst the device fidelity and circuit shallowness are key to the accuracy of the result, we note also that the use of the highest probability bitstring is quite noise resilient, with the limiting factor likely being the ancillae dephasing. Developments in hardware are required to maintain readout fidelity for more readout ancilla, and computational approaches are also available to mitigate the effects of this error~\cite{decoherence_subspace, haghshenas2025digitalquantummagnetismfrontier}.

\section{Discussion \& conclusions}

In this paper, we have reported the use of multi-ancilla quantum phase estimation for the determination of molecular ground-state electronic energies as a benchmarking method for near-term quantum devices.  Our approach is intended to be conceptually simple and low in shot count.  The ground state energy of molecular hydrogen in an STO-3G basis was calculated to $8.9\times 10^{-16}$ hartree when compared against direct diagonalization of the Trotterized unitary, substantially exceeding chemical accuracy and the precision afforded by variational experiments.

Use of our approach as a benchmarking method allowed for elucidation of multiple sources of error.  In Section~\ref{sec:h2}, the impact of Trotter error was observed through a smooth increase in accuracy associated with higher bond lengths; conversely, discretization error was identifiable through its lack of correlation with bond length. In Section~\ref{sec:benchmarking}, we demonstrated the impact of increasing hardware noise, observing the reduced impact of quantization error.  The qualitative differences in the results observed reinforce the efficacy of the method as a benchmark.

Outside a benchmarking context, a clear drawback of our approach is a lack of scalability with regards to system register size.  Although it is possible that alternative unitary decompositions~\cite{krolHighlyEfficientDecomposition2024} could be used to extend beyond two system register qubits, such decompositions inherently scale exponentially with regards to the qubit count. While this provides strong practical limitations on extension beyond two system register qubits, novel decomposition approaches have been shown to be effective in fast-forwarding moderately larger circuits~\cite{wierichsRecursiveCartanDecompositions2025}.  The combination of these techniques with our approach may yield progress in extending system sizes towards the classically intractable regime.  It is further possible that our approach could be useful within a context where fixed-width subcircuits could be found -- for instance, through the application of embedding techniques~\cite{rossmannekQuantumEmbeddingMethod2023, sunQuantumEmbeddingTheories2016} or novel ansatzes.

Due to its purpose as a benchmarking tool, a primary goal of our approach was conceptual simplicity and agnosticity with regards to architecture.  This is reflected in the lack of application of error mitigation, detection or correction techniques.  It is likely that the use of such techniques would improve upon the results presented, potentially allowing for even more precision.  Again for simplicity, all circuit optimization was performed using built-in optimization passes of the \texttt{tket} compiler.  The effectiveness of this reinforces the potential for general-purpose optimization and compilation workflows within an applications context, and the need for the development of a corresponding software stack.

As quantum devices move closer towards application-scale, classically-unsimulable computation, the development of scalable, low shot-count benchmarks will become increasingly important.  Through the use of an algorithm generally intended for fault-tolerant devices, we avoid many of the pitfalls of variational methods.  This emphasizes the importance of the consideration of these algorithms in addition to near-term methods with high shot-count.

The approach described in this paper is reflective of the central appeal of the use of quantum computers for this problem: the ability to obtain highly accurate results, where the classical equivalent may be intractable.  We suggest that benchmarks grounded in this idea -- such as that described in this paper -- will firmly support the development of this ability, as the field enters the early fault-tolerant era.

\begin{acknowledgments}
    The authors thank Regina Finsterhoelzl, Carlo Gaggioli, Yuta Kikuchi and Maria Tudorovskaya for their insightful comments on a draft of this paper.  They also thank Alec Edgington for assistance with the \texttt{tket} compiler's unitary recompilation pass, and the Quantinuum chemistry team for useful discussions.
\end{acknowledgments}

\newpage
\appendix

\section{\ce{H2} phase and energy estimates}
\label{app:h2_phases}

In this appendix we report the phase and energy results obtained in Section~\ref{sec:h2}.  Below $50$ readout qubits, $-\log_2{\Delta E}$ is consistently greater than the number of readout qubits, with the exception of the $9$ readout qubit case.  This is the expected behaviour, as the outcome observed corresponds to the closest possible phase result to the true phase.  The $9$ readout qubit case is discussed in the following appendix.

For $50$ readout qubits and greater, the reported phase error is an exact power of $2$.  This reinforces our hypothesis in Section~\ref{sec:h2} that the error in this case is bound by classical machine precision in data analysis, rather than hardware noise.  It would be possible to repeat the comparison using quadruple precision to confirm this suspicion; however, without extensive modifications to the device software stack, such an analysis would likely result in complex error behaviour.  We leave a study of this behavior for future work.

\begin{longtable}{rrrr}
    \caption{Phases observed using QPE for \ce{H2}.}\\
   \toprule
    Qubits & Phase & Error & log2(Error) \\ \midrule
    \begin{filecontents*}{phases_appendix.dat}
qubits,phase,phaseerror,logphaseerror,energy,energyerror,logenergyerror
1,0.0,0.4159795778272062,-1.2654153927003986,0.0,0.8319591556544124,-0.26541539270039866
2,0.5,0.08402042217279382,-3.5731161549016646,1.0,0.16804084434558764,-2.5731161549016646
3,0.5,0.08402042217279382,-3.5731161549016646,1.0,0.16804084434558764,-2.5731161549016646
4,0.375,0.04097957782720618,-4.608951068034968,0.75,0.08195915565441236,-3.608951068034968
5,0.4375,0.02152042217279382,-5.53814980980668,0.875,0.04304084434558764,-4.53814980980668
6,0.40625,0.009729577827206182,-6.683407077786199,0.8125,0.019459155654412363,-5.683407077786199
7,0.421875,0.005895422172793818,-7.406189155971416,0.84375,0.011790844345587637,-6.406189155971416
8,0.4140625,0.0019170778272061817,-9.026875377679367,0.828125,0.0038341556544123634,-8.026875377679367
9,0.41796875,0.0019891721727938183,-8.973616130619295,0.8359375,0.003978344345587637,-7.973616130619296
10,0.416015625,3.6047172793818305e-05,-14.759754362163092,0.83203125,7.209434558763661e-05,-13.759754362163092
11,0.416015625,3.6047172793818305e-05,-14.759754362163092,0.83203125,7.209434558763661e-05,-13.759754362163092
12,0.416015625,3.6047172793818305e-05,-14.759754362163092,0.83203125,7.209434558763661e-05,-13.759754362163092
13,0.416015625,3.6047172793818305e-05,-14.759754362163092,0.83203125,7.209434558763661e-05,-13.759754362163092
14,0.416015625,3.6047172793818305e-05,-14.759754362163092,0.83203125,7.209434558763661e-05,-13.759754362163092
15,0.41595458984375,2.4987983456181695e-05,-15.288405994586928,0.8319091796875,4.997596691236339e-05,-14.288405994586928
16,0.415985107421875,5.529594668818305e-06,-17.464394837637727,0.83197021484375,1.105918933763661e-05,-16.464394837637727
17,0.415985107421875,5.529594668818305e-06,-17.464394837637727,0.83197021484375,1.105918933763661e-05,-16.464394837637727
18,0.41597747802734375,2.099799862431695e-06,-18.861316742021835,0.8319549560546875,4.19959972486339e-06,-17.861316742021835
19,0.4159812927246094,1.714897403193305e-06,-19.153446302157455,0.8319625854492188,3.42979480638661e-06,-18.153446302157455
20,0.41597938537597656,1.9245122961919492e-07,-22.309003775292727,0.8319587707519531,3.8490245923838984e-07,-21.309003775292727
21,0.41597938537597656,1.9245122961919492e-07,-22.309003775292727,0.8319587707519531,3.8490245923838984e-07,-21.309003775292727
22,0.41597938537597656,1.9245122961919492e-07,-22.309003775292727,0.8319587707519531,3.8490245923838984e-07,-21.309003775292727
23,0.41597962379455566,4.596734948236758e-08,-24.37481527760562,0.8319592475891113,9.193469896473516e-08,-23.37481527760562
24,0.41597962379455566,4.596734948236758e-08,-24.37481527760562,0.8319592475891113,9.193469896473516e-08,-23.37481527760562
25,0.4159795641899109,1.3637295293023044e-08,-26.12786721842221,0.8319591283798218,2.7274590586046088e-08,-25.12786721842221
26,0.4159795641899109,1.3637295293023044e-08,-26.12786721842221,0.8319591283798218,2.7274590586046088e-08,-25.12786721842221
27,0.4159795790910721,1.2638659008246123e-09,-29.559509455723195,0.8319591581821442,2.5277318016492245e-09,-28.559509455723195
28,0.4159795790910721,1.2638659008246123e-09,-29.559509455723195,0.8319591581821442,2.5277318016492245e-09,-28.559509455723195
29,0.4159795790910721,1.2638659008246123e-09,-29.559509455723195,0.8319591581821442,2.5277318016492245e-09,-28.559509455723195
30,0.41597957722842693,5.987792484063448e-10,-30.63725672537313,0.8319591544568539,1.1975584968126896e-09,-29.63725672537313
31,0.4159795781597495,3.3254332620913374e-10,-31.48573863101175,0.831959156319499,6.650866524182675e-10,-30.48573863101175
32,0.4159795776940882,1.3311796109860552e-10,-32.8065757071529,0.8319591553881764,2.6623592219721104e-10,-31.806575707152906
33,0.41597957792691886,9.971268255526411e-11,-33.2234320296498,0.8319591558538377,1.9942536511052822e-10,-32.2234320296498
34,0.41597957781050354,1.6702639271670705e-11,-35.80113295523528,0.8319591556210071,3.340527854334141e-11,-34.80113295523528
35,0.41597957781050354,1.6702639271670705e-11,-35.80113295523528,0.8319591556210071,3.340527854334141e-11,-34.80113295523528
36,0.4159795778396074,1.2401191185062999e-11,-36.23073033974196,0.8319591556792147,2.4802382370125997e-11,-35.23073033974196
37,0.41597957782505546,2.1507240433038533e-12,-38.758314712398374,0.8319591556501109,4.3014480866077065e-12,-37.758314712398374
38,0.41597957782505546,2.1507240433038533e-12,-38.758314712398374,0.8319591556501109,4.3014480866077065e-12,-37.758314712398374
39,0.41597957782869344,1.4872547637878597e-12,-39.290485339293284,0.8319591556573869,2.9745095275757194e-12,-38.290485339293284
40,0.41597957782687445,3.3173463975799677e-13,-41.455035567210764,0.8319591556537489,6.634692795159935e-13,-40.455035567210764
41,0.41597957782687445,3.3173463975799677e-13,-41.455035567210764,0.8319591556537489,6.634692795159935e-13,-40.455035567210764
42,0.4159795778273292,1.2301271112846734e-13,-42.886257833950815,0.8319591556546584,2.460254222569347e-13,-41.886257833950815
43,0.4159795778271018,1.0436096431476471e-13,-43.123483053435,0.8319591556542036,2.0872192862952943e-13,-42.123483053435
44,0.4159795778272155,9.325873406851315e-15,-46.60768257722124,0.831959155654431,1.865174681370263e-14,-45.60768257722124
45,0.4159795778272155,9.325873406851315e-15,-46.60768257722124,0.831959155654431,1.865174681370263e-14,-45.60768257722124
46,0.4159795778272155,9.325873406851315e-15,-46.60768257722124,0.831959155654431,1.865174681370263e-14,-45.60768257722124
47,0.4159795778272013,4.884981308350689e-15,-47.540568381362704,0.8319591556544026,9.769962616701378e-15,-46.540568381362704
48,0.4159795778272084,2.220446049250313e-15,-48.67807190511264,0.8319591556544168,4.440892098500626e-15,-47.67807190511264
49,0.41597957782720485,1.3322676295501878e-15,-49.415037499278846,0.8319591556544097,2.6645352591003757e-15,-48.415037499278846
50,0.4159795778272066,4.440892098500626e-16,-51.0,0.8319591556544133,8.881784197001252e-16,-50.0
51,0.41597957782720574,4.440892098500626e-16,-51.0,0.8319591556544115,8.881784197001252e-16,-50.0
52,0.41597957782720574,4.440892098500626e-16,-51.0,0.8319591556544115,8.881784197001252e-16,-50.0
53,0.41597957782720596,2.220446049250313e-16,-52.0,0.8319591556544119,4.440892098500626e-16,-51.0
54,0.41597957782720596,2.220446049250313e-16,-52.0,0.8319591556544119,4.440892098500626e-16,-51.0
55,0.41597957782720596,2.220446049250313e-16,-52.0,0.8319591556544119,4.440892098500626e-16,-51.0
\end{filecontents*}
\csvreader[
head to column names
]{phases_appendix.dat}{}{%
\qubits\ & \phase & \phaseerror & \logphaseerror  \\
} \\ \bottomrule
    \end{longtable}

\begin{longtable}{rrrr}
    \caption{Energetics observed using QPE for \ce{H2}.} \\
   \toprule
    Qubits & Energy & Error & log2(error) \\ \midrule
    \csvreader[
head to column names
]{phases_appendix.dat}{}{%
\qubits\ & \energy & \energyerror & \logenergyerror \\
} \\ \bottomrule
    \end{longtable}
\newpage
\section{Precision and discretization effects}
\label{app:shots}
In this appendix, we consider several examples of observed shot counts in various experiments described in the main text.  As discretization error results in an unpredictable distribution of measurement outcomes even in the absence of noise, the experimental outcomes are compared against noiseless emulation  Deviations from the noiseless result are considered attributable to error.

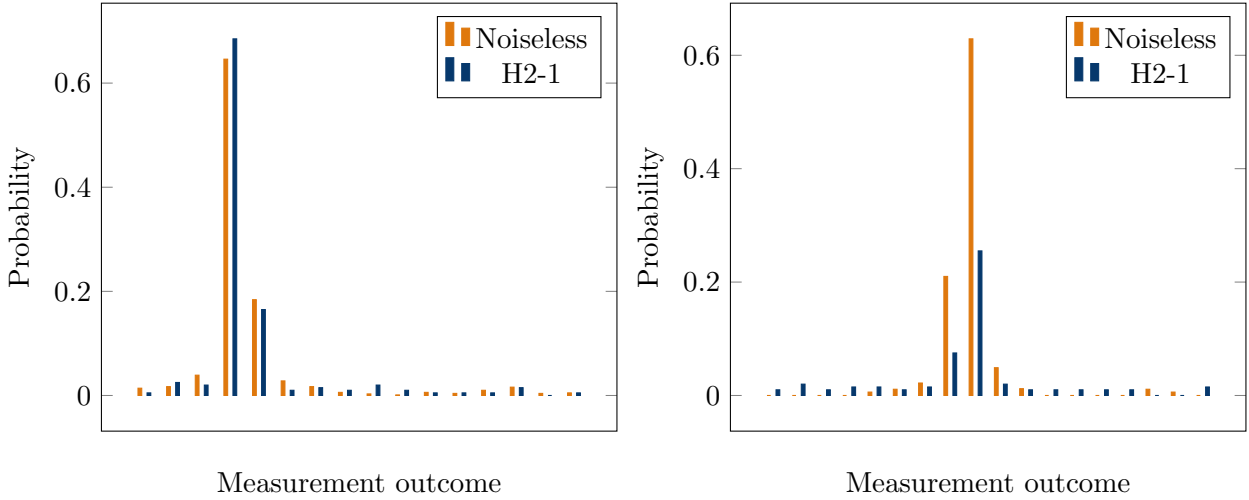
\begin{figure}
\begin{center}
    \begin{tabular}{rl}
 \begin{tikzpicture}[baseline, trim axis left]
     \pgfplotstableread[col sep = comma]{app_shots_4.dat}{\fourdat}
       \pgfplotstableread[col sep = comma]{app_shots_31.dat}{\thirtyonedat}

  \begin{axis}[
  ybar,
  xlabel={Measurement outcome},
  ylabel={Probability},
  legend pos = north east,
    cycle list name=color list,
  bar width = 0.05cm,
  xtick style = {draw=none},
  xticklabel=\empty,
   ]
  \addplot+[color=dualAnalagous2,fill=dualAnalagous2]
   table[x={index}, y={nl}]
    {\fourdat};
   \addplot+[color=dualAnalagous1,fill=dualAnalagous1]
  table[x={index}, y={hw}]
    {\fourdat};
        \legend{Noiseless,H2-1}
  \end{axis}
 \end{tikzpicture}
        &
         \begin{tikzpicture}[baseline, trim axis right]
  \pgfplotstableread[col sep = comma]{app_shots_31.dat}{\thirtyonedat}
  \begin{axis}[
  ybar,
  xlabel={Measurement outcome},
  ylabel={Probability},
  legend pos = north east,
    cycle list name=color list,
  bar width = 0.05cm,
      xtick style = {draw=none},
  xticklabel=\empty,
   ]
  \addplot+[color=dualAnalagous2,fill=dualAnalagous2]
   table[x={index},y={nl}]
    {\thirtyonedat};
   \addplot+[color=dualAnalagous1,fill=dualAnalagous1]
  table[x={index},y={hw}]
    {\thirtyonedat};
        \legend{Noiseless,H2-1}
  \end{axis}
             \end{tikzpicture}
        \\
\end{tabular}
\end{center}
    \caption{A comparison of measurement outcome distributions for $4$ (left) and $31$ (right) readout qubits.  Due to greater hardware noise from deeper circuits, increasing the number of readout qubits causes the obtained shot distribution to broaden, with a greater difference between hardware results and noiseless emulation.  Low-probability outcomes are omitted on the right-hand plot for clarity.}
    \label{fig:appshots}
    \end{figure}

We first consider the impact of increasing readout qubit count upon the measurement outcomes.  As shown in the left of Figure~\ref{fig:appshots}, at a low readout qubit count we obtain a measurement distribution that is extremely close to the expected distribution.  Conversely, as we increase to $31$ readout qubits (right of Figure~\ref{fig:appshots}), the distribution of obtained results is much more dispersed, with the central peak greatly reduced in probability from the expected distribution.  This is attributable to the deeper circuits requiring more gates and longer computation times, increasing the effect of gate error and decoherence effects.

\begin{figure}
\begin{center}
    \begin{tabular}{rl}
 \begin{tikzpicture}[baseline, trim axis left]
     \pgfplotstableread[col sep = comma]{app_shots_9.dat}{\ninedat}
  \begin{axis}[
  ybar,
  xlabel={Measurement outcome},
  ylabel={Probability},
  legend pos = north east,
    cycle list name=color list,
  bar width = 0.05cm,
    xtick style = {draw=none},
  xticklabel=\empty,
   ]
  \addplot+[color=dualAnalagous2,fill=dualAnalagous2]
   table[x={index}, y={nl}]
    {\ninedat};
   \addplot+[color=dualAnalagous1,fill=dualAnalagous1]
  table[x={index}, y={hw}]
    {\ninedat};
        \legend{Noiseless,H2-1}
  \end{axis}
 \end{tikzpicture}
        &
         \begin{tikzpicture}[baseline, trim axis right]
  \pgfplotstableread[col sep = comma]{app_shots_broken.dat}{\brokendat}
  \begin{axis}[
  ybar,
  xlabel={Measurement outcome},
  ylabel={Probability},
  legend pos = north east,
    cycle list name=color list,
  bar width = 0.05cm,
    xtick style = {draw=none},
  xtick={\empty},
   ]
  \addplot+[color=dualAnalagous2,fill=dualAnalagous2]
   table[x={index},y={prob1}]
    {\brokendat};
   \addplot+[color=dualAnalagous1,fill=dualAnalagous1]
  table[x={index},y={prob2}]
    {\brokendat};
        \legend{Initial experiment,Repeated experiment}
  \end{axis}
             \end{tikzpicture}
        \\
\end{tabular}
\end{center}
    \caption{Shot results of anomalous experiments.  Low-probability outcomes are excluded for clarity.  Left: $9$ readout qubits at equilibrium geometry.  The peak is reversed from the noiseless emulated result, attributable to high discretization error and the limited number of shots.  Right: the anomalous stretched result discussed in the main text.  For the initial experiment, the wrong outcome is obtained.  A repeated experiment obtained the correct result, albeit with low probability.}
    \label{fig:appbroken}
    \end{figure}
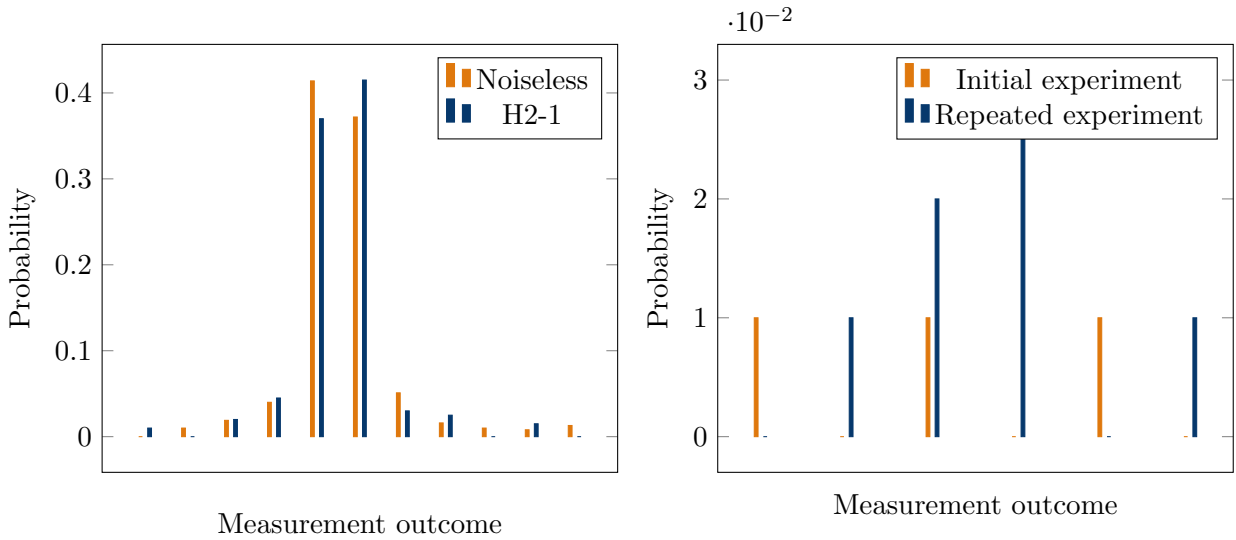

In Figure~\ref{fig:appbroken}, we consider two anomalous results noted above.  The left plot shows the $9$ readout qubit case of \ce{H2} in STO-3G at equilibrium.  In Appendix~\ref{app:h2_phases}), it was discussed that less than $9$ bits of precision were obtained.  Figure~\ref{fig:appbroken} shows that this result was due to the fact that the expected peak has a high probability of obtaining two possible results, as the true phase is close to equidistant between their corresponding phases.  Examining the experimental results in this plot, it is evident that they maintain this behavior -- the two expected results have high probability -- but the relative height of the peaks is reversed, due to statistical chance induced by the limited number of shots.  As our phase estimate is obtained by simply picking the most likely possible outcome, in this case we obtain the incorrect peak, corresponding to a phase with the least significant bit flipped.  The phase is therefore accurate to $8$ bits of precision, rather than $9$.  This result is not unexpected: in QPE, discretization error inevitably results in a chance of failure in this manner, motivating our cited precision as being one bit less than the amount actually read from the device.  The $9$ readout qubit case does indeed fall within this quoted precision.

On the right-hand side of Figure~\ref{fig:appbroken}, we consider the outlier result in Figure~\ref{fig:dissocation} of the main text, with $50$ readout qubits and $10^4$ Trotter steps at a stretched geometry.  This result obtained a substantially larger error than similar experiments reported in Section~\ref{sec:h2}.  While this amount of qubits is beyond the capabilities of the emulator, we can compare against a subsequent hardware experiment.  The right-hand plot of Figure~\ref{fig:appbroken} shows that the repeated experiment obtained the correct phase with reasonably high probability.  Conversely, the initial experiment was particularly noisy, resulting in other outcomes obtaining equal probability, with one of these outcomes selected as the most likely outcome.

\clearpage
\section{All error scaling}
These error model calculations (see ~\cite{SystemModelH2} for error model details) of the high-precision phase estimation algorithm utilized a single Trotter step for the time evolution operator, Hartree-Fock as the initial state, and are at equilibrium bond length. See the main text for further details. A value of 10 represents making all model errors ``10 times worse'', for example by increasing the probability for a 2-qubit gate error by a factor of 10 as well as increasing the rate at which qubit dephasing occurs, among other factors (measurement error, crosstalk, etc). Conversely, an error scale below 1 represents a less noisy device than the H2 system, across all error model parameters. Through summing the three most likely phases, we remove some discretization effects (in contrast to Figure\ref{all_error_scaled} in the main text), and so the metric more smoothly decays due to noise.
\begin{figure}[h]
    \centering
\begin{tikzpicture}[scale=0.8]

\definecolor{cornflowerblue85170255}{RGB}{85,170,255}
\definecolor{cyan}{RGB}{0,255,255}
\definecolor{darkgray176}{RGB}{176,176,176}
\definecolor{deepskyblue42213255}{RGB}{42,213,255}
\definecolor{magenta}{RGB}{255,0,255}
\definecolor{magenta21341255}{RGB}{213,41,255}
\definecolor{mediumorchid17085255}{RGB}{170,85,255}
\definecolor{mediumslateblue128127255}{RGB}{128,127,255}

\begin{axis}[
tick align=center,
tick pos=both,
legend pos = outer north east,
title={H$_2$ all errors scaled},
x grid style={darkgray176},
xlabel={Readout qubits},
xmin=0, xmax=30,
xtick style={color=black},
xtick={0,10,20,30},
xticklabels={
  \(\displaystyle {0}\),
  \(\displaystyle {10}\),
  \(\displaystyle {20}\),
  \(\displaystyle {30}\)
},
y grid style={darkgray176},
ylabel={P\(\displaystyle _{sum\, of\, 3}\)},
ymin=-0.05, ymax=1.05,
ytick style={color=black},
ytick={-0.2,0,0.2,0.4,0.6,0.8,1,1.2},
yticklabels={
  \(\displaystyle {\ensuremath{-}0.2}\),
  \(\displaystyle {0.0}\),
  \(\displaystyle {0.2}\),
  \(\displaystyle {0.4}\),
  \(\displaystyle {0.6}\),
  \(\displaystyle {0.8}\),
  \(\displaystyle {1.0}\),
  \(\displaystyle {1.2}\)
}
]
\addplot [draw=deepskyblue42213255, fill=deepskyblue42213255, mark=x,mark size=4pt, only marks]
table{%
x  y
30 0.735
29 0.675
28 0.83
27 0.885
26 0.77
25 0.83
24 0.79
23 0.855
22 0.775
21 0.89
20 0.91
19 0.81
18 0.885
17 0.875
16 0.965
15 0.835
14 0.87
13 0.915
12 0.985
11 0.995
10 0.99
nan nan
8 0.945
7 0.89
6 0.91
5 0.885
4 0.905
3 0.91
2 0.97
1 1
};
\addplot [draw=cornflowerblue85170255, fill=cornflowerblue85170255, mark=x,mark size=4pt, only marks]
table{%
x  y
30 0.57
29 0.51
28 0.62
27 0.68
26 0.77
25 0.675
24 0.7
23 0.77
22 0.775
21 0.855
20 0.87
19 0.785
18 0.845
17 0.825
16 0.895
15 0.875
14 0.845
13 0.895
12 0.98
11 0.975
10 0.975
nan nan
8 0.9
7 0.845
6 0.85
5 0.915
4 0.905
3 0.88
2 0.97
1 1
};
\addplot [draw=mediumslateblue128127255, fill=mediumslateblue128127255, mark=x,mark size=4pt, only marks]
table{%
x  y
30 0.295
29 0.35
28 0.405
27 0.435
26 0.475
25 0.535
24 0.53
23 0.645
22 0.575
21 0.78
20 0.785
19 0.74
18 0.805
17 0.79
16 0.815
15 0.78
14 0.825
13 0.915
12 0.92
11 0.945
10 0.955
9 0.815
8 0.885
7 0.82
6 0.875
5 0.865
4 0.835
3 0.92
2 0.975
1 1
};
\addplot [draw=mediumorchid17085255, fill=mediumorchid17085255, mark=x,mark size=4pt, only marks]
table{%
x  y
30 0.0800000000000001
29 0.11
28 0.145
27 0.265
26 0.295
25 0.3
24 0.35
23 0.475
22 0.415
21 0.625
20 0.705
19 0.575
18 0.61
17 0.68
16 0.735
15 0.675
14 0.775
13 0.855
12 0.88
11 0.915
10 0.96
nan nan
8 0.85
7 0.83
6 0.85
5 0.84
4 0.895
3 0.89
2 0.98
1 1
};
\addplot [draw=magenta21341255, fill=magenta21341255, mark=x,mark size=4pt, only marks]
table{%
x  y
27 0.0449999999999999
26 0.0449999999999999
25 0.0649999999999999
24 0.095
23 0.095
22 0.21
21 0.295
20 0.405
19 0.355
18 0.395
17 0.51
16 0.55
15 0.49
14 0.535
13 0.665
12 0.715
11 0.74
10 0.84
nan nan
8 0.735
7 0.76
6 0.825
5 0.775
4 0.845
3 0.855
2 0.965
1 1
};
\addplot [draw=magenta, fill=magenta, mark=x,mark size=4pt, only marks]
table{%
x  y
24 0.0349999999999999
23 0.04
nan nan
21 0.105
20 0.145
19 0.145
18 0.175
17 0.225
16 0.335
15 0.315
14 0.35
13 0.47
12 0.45
11 0.615
10 0.66
9 0.545
8 0.73
7 0.67
6 0.7
5 0.76
4 0.815
3 0.885
2 0.965
1 1
};
\addplot [thick, neonorange]
table {%
30 0.894999980926514
29 0.860000014305115
28 0.940000057220459
27 0.995000004768372
26 0.865000009536743
25 0.920000076293945
24 0.889999985694885
23 0.920000076293945
22 0.845000028610229
21 0.930000066757202
20 0.980000019073486
19 0.815000057220459
18 0.889999985694885
17 0.914999961853027
16 0.950000047683716
15 0.879999995231628
14 0.884999990463257
13 0.975000023841858
12 0.980000019073486
11 1
10 1
9 0.855000019073486
8 0.920000076293945
7 0.855000019073486
6 0.899999976158142
5 0.865000009536743
4 0.914999961853027
3 0.879999995231628
2 0.970000028610229
1 1
};
\addplot [semithick, deepskyblue42213255]
table {%
30 0.735000014305115
29 0.674999952316284
28 0.829999923706055
27 0.884999990463257
26 0.769999980926514
25 0.829999923706055
24 0.789999961853027
23 0.855000019073486
22 0.774999976158142
21 0.889999985694885
20 0.910000085830688
19 0.809999942779541
18 0.884999990463257
17 0.875
16 0.964999914169312
15 0.835000038146973
14 0.870000004768372
13 0.914999961853027
12 0.985000014305115
11 0.995000004768372
10 0.990000009536743
9 0.860000014305115
8 0.944999933242798
7 0.889999985694885
6 0.910000085830688
5 0.884999990463257
4 0.904999971389771
3 0.910000085830688
2 0.970000028610229
1 1
};
\addplot [semithick, cornflowerblue85170255]
table {%
30 0.569999933242798
29 0.509999990463257
28 0.620000004768372
27 0.680000066757202
26 0.769999980926514
25 0.674999952316284
24 0.700000047683716
23 0.769999980926514
22 0.774999976158142
21 0.855000019073486
20 0.870000004768372
19 0.785000085830688
18 0.845000028610229
17 0.825000047683716
16 0.894999980926514
15 0.875
14 0.845000028610229
13 0.894999980926514
12 0.980000019073486
11 0.975000023841858
10 0.975000023841858
9 0.829999923706055
8 0.899999976158142
7 0.845000028610229
6 0.850000023841858
5 0.914999961853027
4 0.904999971389771
3 0.879999995231628
2 0.970000028610229
1 1
};
\addplot [semithick, mediumslateblue128127255]
table {%
30 0.294999957084656
28 0.404999971389771
27 0.434999942779541
26 0.475000023841858
25 0.535000085830688
24 0.529999971389771
23 0.644999980926514
22 0.575000047683716
21 0.779999971389771
20 0.785000085830688
19 0.740000009536743
18 0.805000066757202
17 0.789999961853027
16 0.815000057220459
15 0.779999971389771
14 0.825000047683716
13 0.914999961853027
12 0.920000076293945
11 0.944999933242798
10 0.954999923706055
9 0.815000057220459
8 0.884999990463257
7 0.819999933242798
6 0.875
5 0.865000009536743
4 0.835000038146973
3 0.920000076293945
2 0.975000023841858
1 1
};
\addplot [semithick, mediumorchid17085255]
table {%
30 0.0800000429153442
29 0.110000014305115
28 0.144999980926514
27 0.264999985694885
26 0.294999957084656
25 0.299999952316284
24 0.350000023841858
23 0.475000023841858
22 0.414999961853027
21 0.625
20 0.704999923706055
19 0.575000047683716
18 0.610000014305115
17 0.680000066757202
16 0.735000014305115
15 0.674999952316284
14 0.774999976158142
13 0.855000019073486
12 0.879999995231628
11 0.914999961853027
10 0.960000038146973
9 0.759999990463257
8 0.850000023841858
7 0.829999923706055
6 0.850000023841858
5 0.839999914169312
4 0.894999980926514
3 0.889999985694885
2 0.980000019073486
1 1
};
\addplot [semithick, magenta21341255]
table {%
30 0.0149999856948853
29 0.0199999809265137
28 0.0199999809265137
27 0.0449999570846558
26 0.0449999570846558
25 0.065000057220459
24 0.0950000286102295
23 0.0950000286102295
22 0.210000038146973
21 0.294999957084656
20 0.404999971389771
19 0.355000019073486
18 0.394999980926514
17 0.509999990463257
16 0.549999952316284
15 0.490000009536743
14 0.535000085830688
13 0.664999961853027
12 0.714999914169312
11 0.740000009536743
10 0.839999914169312
9 0.730000019073486
8 0.735000014305115
7 0.759999990463257
6 0.825000047683716
5 0.774999976158142
4 0.845000028610229
3 0.855000019073486
2 0.964999914169312
1 1
};
\addplot [semithick, magenta]
table {%
30 0.0149999856948853
27 0.0149999856948853
26 0.0199999809265137
25 0.0199999809265137
24 0.0349999666213989
23 0.0399999618530273
22 0.0549999475479126
21 0.105000019073486
20 0.144999980926514
19 0.144999980926514
18 0.174999952316284
17 0.225000023841858
16 0.335000038146973
15 0.315000057220459
14 0.350000023841858
13 0.470000028610229
12 0.450000047683716
11 0.615000009536743
10 0.660000085830688
9 0.545000076293945
8 0.730000019073486
7 0.670000076293945
6 0.700000047683716
5 0.759999990463257
4 0.815000057220459
3 0.884999990463257
2 0.964999914169312
1 1
};
\legend{ , , , , , ,Noiseless,0.2, 0.5, 1, 2,5,10}

\end{axis}

\end{tikzpicture}
\begin{tikzpicture}[scale=0.8]

\definecolor{cornflowerblue85170255}{RGB}{85,170,255}
\definecolor{cyan}{RGB}{0,255,255}
\definecolor{darkgray176}{RGB}{176,176,176}
\definecolor{deepskyblue42213255}{RGB}{42,213,255}
\definecolor{magenta}{RGB}{255,0,255}
\definecolor{magenta21341255}{RGB}{213,41,255}
\definecolor{mediumorchid17085255}{RGB}{170,85,255}
\definecolor{mediumslateblue128127255}{RGB}{128,127,255}

\begin{axis}[
tick align=center,
tick pos=both,
title={HeH$^+$ all errors scaled},
x grid style={darkgray176},
xlabel={Readout qubits},
xmin=0, xmax=30,
xtick style={color=black},
xtick={0,10,20,30},
xticklabels={
  \(\displaystyle {0}\),
  \(\displaystyle {10}\),
  \(\displaystyle {20}\),
  \(\displaystyle {30}\)
},
y grid style={darkgray176},
ylabel={P\(\displaystyle _{sum\, of\, 3}\)},
ymin=-0.05, ymax=1.05,
ytick style={color=black},
ytick={-0.2,0,0.2,0.4,0.6,0.8,1,1.2},
yticklabels={
  \(\displaystyle {\ensuremath{-}0.2}\),
  \(\displaystyle {0.0}\),
  \(\displaystyle {0.2}\),
  \(\displaystyle {0.4}\),
  \(\displaystyle {0.6}\),
  \(\displaystyle {0.8}\),
  \(\displaystyle {1.0}\),
  \(\displaystyle {1.2}\)
}
]
\addplot [draw=deepskyblue42213255, fill=deepskyblue42213255, mark=x,mark size=4pt, only marks]
table{%
x  y
30 0.105
29 0.165
28 0.195
27 0.185
26 0.21
25 0.135
24 0.375
23 0.235
22 0.565
21 0.455
20 0.525
19 0.51
18 0.655
17 0.695
16 0.63
15 0.735
14 0.7
13 0.835
12 0.76
11 0.86
10 0.9
9 0.83
8 0.88
7 0.95
6 0.98
5 0.815
4 0.91
3 0.96
2 0.99
1 1
};
\addplot [draw=cornflowerblue85170255, fill=cornflowerblue85170255, mark=x,mark size=4pt, only marks]
table{%
x  y
28 0.05
nan nan
26 0.21
nan nan
24 0.095
23 0.075
22 0.335
21 0.175
20 0.235
19 0.22
18 0.415
17 0.47
16 0.445
15 0.625
14 0.61
13 0.65
12 0.695
11 0.74
10 0.875
9 0.755
8 0.83
7 0.935
6 0.915
5 0.825
4 0.93
3 0.965
2 0.99
1 1
};
\addplot [draw=mediumslateblue128127255, fill=mediumslateblue128127255, mark=x,mark size=4pt, only marks]
table{%
x  y
21 0.0600000000000001
nan nan
19 0.0900000000000001
18 0.2
17 0.285
nan nan
15 0.465
14 0.44
13 0.575
12 0.535
11 0.675
10 0.735
9 0.705
8 0.745
7 0.86
6 0.905
nan nan
4 0.925
3 0.975
2 0.995
1 1
};
\addplot [draw=mediumorchid17085255, fill=mediumorchid17085255, mark=x,mark size=4pt, only marks]
table{%
x  y
16 0.085
15 0.205
14 0.305
13 0.35
12 0.415
11 0.445
10 0.49
9 0.605
8 0.645
7 0.745
6 0.825
5 0.735
4 0.89
3 0.95
2 0.985
1 1
};
\addplot [draw=magenta21341255, fill=magenta21341255, mark=x,mark size=4pt, only marks]
table{%
x  y
15 0.0449999999999999
nan nan
13 0.085
12 0.14
11 0.215
10 0.31
nan nan
8 0.425
7 0.535
6 0.64
5 0.545
4 0.785
3 0.845
2 0.965
1 1
};
\addplot [draw=magenta, fill=magenta, mark=x,mark size=4pt, only marks]
table{%
x  y
11 0.0549999999999999
10 0.0800000000000001
nan nan
8 0.18
7 0.305
6 0.345
5 0.475
4 0.67
3 0.795
2 0.92
1 1
};
\addplot [thick, neonorange]
table {%
30 0.850000023841858
29 0.920000076293945
28 0.970000028610229
27 0.835000038146973
26 0.914999961853027
25 0.904999971389771
24 0.920000076293945
23 0.855000019073486
22 0.910000085830688
21 0.865000009536743
20 0.950000047683716
19 0.855000019073486
18 0.910000085830688
17 0.954999923706055
16 0.879999995231628
15 0.910000085830688
14 0.860000014305115
12 0.899999976158142
11 0.904999971389771
10 0.970000028610229
9 0.879999995231628
8 0.964999914169312
7 0.980000019073486
6 0.975000023841858
5 0.894999980926514
4 0.924999952316284
3 0.964999914169312
2 1
1 1
};
\addplot [semithick, deepskyblue42213255]
table {%
30 0.105000019073486
29 0.164999961853027
28 0.194999933242798
27 0.184999942779541
26 0.210000038146973
25 0.134999990463257
24 0.375
23 0.235000014305115
22 0.565000057220459
21 0.455000042915344
20 0.524999976158142
19 0.509999990463257
18 0.654999971389771
17 0.694999933242798
16 0.629999995231628
15 0.735000014305115
14 0.700000047683716
13 0.835000038146973
12 0.759999990463257
11 0.860000014305115
10 0.899999976158142
9 0.829999923706055
8 0.879999995231628
7 0.950000047683716
6 0.980000019073486
5 0.815000057220459
4 0.910000085830688
3 0.960000038146973
2 0.990000009536743
1 1
};
\addplot [semithick, cornflowerblue85170255]
table {%
30 0.0249999761581421
29 0.0399999618530273
28 0.0499999523162842
27 0.0349999666213989
26 0.210000038146973
25 0.0549999475479126
24 0.0950000286102295
23 0.0750000476837158
22 0.335000038146973
21 0.174999952316284
20 0.235000014305115
19 0.220000028610229
18 0.414999961853027
17 0.470000028610229
16 0.444999933242798
15 0.625
14 0.610000014305115
13 0.649999976158142
11 0.740000009536743
10 0.875
9 0.754999995231628
8 0.829999923706055
7 0.934999942779541
6 0.914999961853027
5 0.825000047683716
4 0.930000066757202
3 0.964999914169312
2 0.990000009536743
1 1
};
\addplot [semithick, mediumslateblue128127255]
table {%
30 0.0199999809265137
29 0.0149999856948853
28 0.0299999713897705
27 0.0249999761581421
26 0.0149999856948853
25 0.0249999761581421
24 0.0449999570846558
23 0.0399999618530273
22 0.110000014305115
21 0.059999942779541
20 0.0700000524520874
19 0.0900000333786011
18 0.200000047683716
17 0.284999966621399
16 0.225000023841858
15 0.465000033378601
14 0.440000057220459
13 0.575000047683716
12 0.535000085830688
11 0.674999952316284
10 0.735000014305115
9 0.704999923706055
8 0.745000004768372
7 0.860000014305115
6 0.904999971389771
5 0.815000057220459
4 0.924999952316284
3 0.975000023841858
2 0.995000004768372
1 1
};
\addplot [semithick, mediumorchid17085255]
table {%
30 0.0149999856948853
25 0.0149999856948853
24 0.0349999666213989
23 0.0149999856948853
22 0.0349999666213989
21 0.0299999713897705
19 0.0399999618530273
18 0.059999942779541
17 0.0850000381469727
16 0.0850000381469727
15 0.205000042915344
14 0.305000066757202
13 0.350000023841858
12 0.414999961853027
11 0.444999933242798
10 0.490000009536743
9 0.605000019073486
8 0.644999980926514
7 0.745000004768372
6 0.825000047683716
5 0.735000014305115
4 0.889999985694885
3 0.950000047683716
2 0.985000014305115
1 1
};
\addplot [semithick, magenta21341255]
table {%
30 0.0149999856948853
23 0.0149999856948853
22 0.0199999809265137
21 0.0199999809265137
20 0.0149999856948853
19 0.0199999809265137
18 0.0199999809265137
17 0.0299999713897705
16 0.0349999666213989
14 0.0549999475479126
13 0.0850000381469727
12 0.139999985694885
11 0.215000033378601
10 0.309999942779541
9 0.294999957084656
8 0.424999952316284
7 0.535000085830688
6 0.639999985694885
5 0.545000076293945
4 0.785000085830688
3 0.845000028610229
2 0.964999914169312
1 1
};
\addplot [semithick, magenta]
table {%
30 0.0149999856948853
19 0.0149999856948853
18 0.0199999809265137
15 0.0199999809265137
13 0.0299999713897705
12 0.0399999618530273
11 0.0549999475479126
10 0.0800000429153442
9 0.125
8 0.180000066757202
7 0.305000066757202
6 0.345000028610229
5 0.475000023841858
4 0.670000076293945
2 0.920000076293945
1 1
};
\end{axis}

\end{tikzpicture}
    \caption{Probability to read the three most likely phases for between 1 and 30 readout ancillae as a function of the H2 ion-trap emulator error model scaling ($\times 0$ (noiseless) at top in blue $\rightarrow$ $\times 10$ is bottom in pink) . `$\times$' marks denote results where the highest probability phase of the noisy result matches the highest probability phase of the noiseless result for 200 shots. }
    \label{allscaled_p3max}
\end{figure}
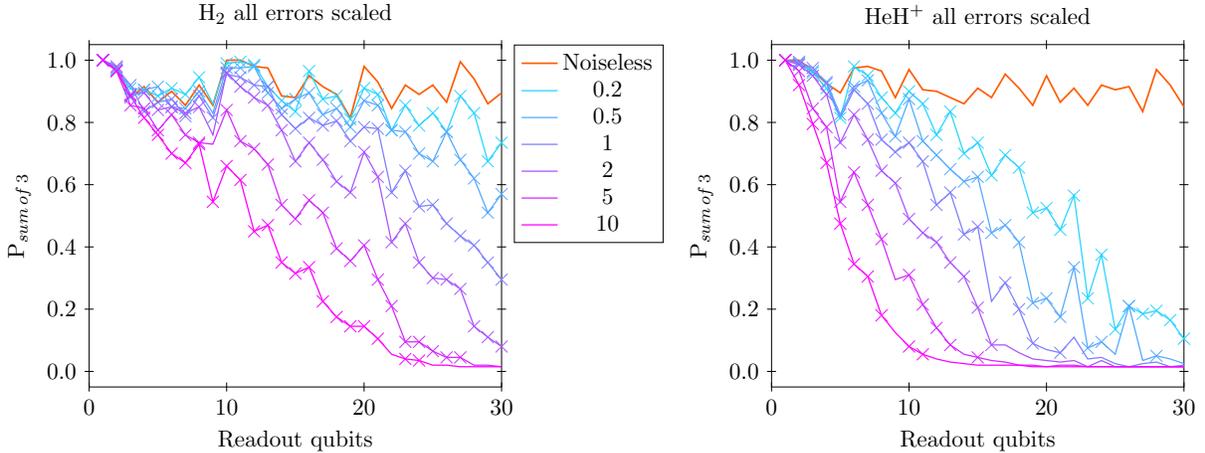

\section{Specific error model parameter scaling }
As two-qubit gate error probabilities are typically orders of magnitude larger than one-qubit (or other) errors, we expect them to dominate. In our circuits we also expect a significant effect from memory error. This is because, for example, the ancilla which is the control for $U^{2^{0}}$ will be left idling whilst all other successive controls are applied. This allows time for coherent and incoherent dephasing effects to occur. We also expect this effect to be exacerbated for larger numbers of qubits as more ion-transport time will be required to bring ions to- and from- the active gate zone. Those results show that even for large error model scalings ($\times 100$), the device is performant enough, and the circuit shallow enough, that the QPE algorithm succeeds for \ce{H2} for more than 12 readout qubits; approximately enough for \lq chemical accuracy\rq. However, for \ce{HeH+} the deeper circuits demonstrate greater noise model sensitivity, with a more pronounced drop in $P_{max}$ and with more incorrect phases being obtained. Tentatively, we suggest these results are more sensitive to memory than two-qubit gate errors as incorrect readout is obtained with higher probability for lower numbers of readout qubits. These error model calculations utilized a single Trotter step for the time evolution operator, Hartree-Fock as the initial state, and are at equilibrium bond length.

\subsection{Memory error scaling}
\begin{figure}[h]
    \centering
\begin{tikzpicture}[scale=0.8]

\definecolor{cornflowerblue109146255}{RGB}{109,146,255}
\definecolor{cornflowerblue73182255}{RGB}{73,182,255}
\definecolor{cyan}{RGB}{0,255,255}
\definecolor{darkgray176}{RGB}{176,176,176}
\definecolor{deepskyblue36219255}{RGB}{36,219,255}
\definecolor{magenta}{RGB}{255,0,255}
\definecolor{magenta21936255}{RGB}{219,36,255}
\definecolor{mediumorchid18273255}{RGB}{182,73,255}
\definecolor{mediumslateblue146109255}{RGB}{146,109,255}

\begin{axis}[
tick align=center,
tick pos=both,
legend pos = outer north east,
title={H$_2$ memory error scaled},
x grid style={darkgray176},
xlabel={Readout qubits},
xmin=0, xmax=20,
xtick style={color=black},
xtick={0,5,10,15,20},
xticklabels={
  \(\displaystyle {0}\),
  \(\displaystyle {5}\),
  \(\displaystyle {10}\),
  \(\displaystyle {15}\),
  \(\displaystyle {20}\)
},
y grid style={darkgray176},
ylabel={P\(\displaystyle _{max}\)},
ymin=-0.05, ymax=1.05,
ytick style={color=black},
ytick={-0.2,0,0.2,0.4,0.6,0.8,1,1.2},
yticklabels={
  \(\displaystyle {\ensuremath{-}0.2}\),
  \(\displaystyle {0.0}\),
  \(\displaystyle {0.2}\),
  \(\displaystyle {0.4}\),
  \(\displaystyle {0.6}\),
  \(\displaystyle {0.8}\),
  \(\displaystyle {1.0}\),
  \(\displaystyle {1.2}\)
}
]
\addplot [draw=deepskyblue36219255, fill=deepskyblue36219255, mark=x,mark size=4pt, only marks]
table{%
x  y
18 0.72
17 0.585
16 0.8
15 0.47
14 0.665
13 0.855
12 0.9
11 0.935
10 0.94
9 0.38
8 0.7
7 0.585
6 0.705
5 0.65
4 0.625
3 0.645
2 0.93
1 0.605
};
\addplot [draw=cornflowerblue73182255, fill=cornflowerblue73182255, mark=x,mark size=4pt, only marks]
table{%
x  y
18 0.645
17 0.555
16 0.755
15 0.52
14 0.645
13 0.895
12 0.905
11 0.925
10 0.9
nan nan
8 0.82
7 0.53
6 0.68
5 0.675
4 0.7
3 0.66
2 0.925
1 0.69
};
\addplot [draw=cornflowerblue109146255, fill=cornflowerblue109146255, mark=x,mark size=4pt, only marks]
table{%
x  y
18 0.68
17 0.575
16 0.78
15 0.475
14 0.7
13 0.825
12 0.91
11 0.935
10 0.89
9 0.41
8 0.775
7 0.515
6 0.61
5 0.635
4 0.64
3 0.675
2 0.875
1 0.62
};
\addplot [draw=mediumslateblue146109255, fill=mediumslateblue146109255, mark=x,mark size=4pt, only marks]
table{%
x  y
18 0.615
17 0.505
16 0.76
15 0.45
14 0.675
13 0.815
12 0.915
11 0.88
10 0.92
9 0.39
8 0.81
7 0.56
6 0.71
5 0.64
4 0.685
3 0.69
2 0.88
1 0.695
};
\addplot [draw=mediumorchid18273255, fill=mediumorchid18273255, mark=x,mark size=4pt, only marks]
table{%
x  y
18 0.5
17 0.5
16 0.645
15 0.485
14 0.585
13 0.795
12 0.84
11 0.88
10 0.92
nan nan
8 0.76
7 0.55
6 0.66
5 0.585
4 0.625
3 0.665
2 0.875
1 0.59
};
\addplot [draw=magenta21936255, fill=magenta21936255, mark=x,mark size=4pt, only marks]
table{%
x  y
18 0.45
17 0.395
16 0.625
15 0.415
14 0.545
13 0.725
12 0.82
11 0.835
10 0.91
nan nan
8 0.695
7 0.6
6 0.7
5 0.71
4 0.65
3 0.605
2 0.925
1 0.585
};
\addplot [draw=magenta, fill=magenta, mark=x,mark size=4pt, only marks]
table{%
x  y
15 0.0449999999999999
14 0.16
13 0.185
12 0.3
11 0.56
10 0.615
9 0.27
8 0.61
7 0.485
6 0.62
5 0.55
4 0.58
3 0.625
2 0.855
1 0.6
};
\addplot [semithick, cyan]
table {%
18 0.649999976158142
17 0.615000009536743
16 0.764999985694885
15 0.509999990463257
14 0.745000004768372
13 0.795000076293945
12 0.950000047683716
10 0.940000057220459
9 0.419999957084656
8 0.789999961853027
7 0.565000057220459
6 0.700000047683716
5 0.579999923706055
4 0.670000076293945
3 0.664999961853027
2 0.910000085830688
1 0.664999961853027
};
\addplot [semithick, deepskyblue36219255]
table {%
18 0.720000028610229
17 0.585000038146973
16 0.799999952316284
15 0.470000028610229
14 0.664999961853027
13 0.855000019073486
12 0.899999976158142
11 0.934999942779541
10 0.940000057220459
9 0.379999995231628
8 0.700000047683716
7 0.585000038146973
6 0.704999923706055
5 0.649999976158142
4 0.625
3 0.644999980926514
2 0.930000066757202
1 0.605000019073486
};
\addplot [semithick, cornflowerblue73182255]
table {%
18 0.644999980926514
17 0.555000066757202
16 0.754999995231628
15 0.519999980926514
14 0.644999980926514
13 0.894999980926514
12 0.904999971389771
11 0.924999952316284
10 0.899999976158142
9 0.375
8 0.819999933242798
7 0.529999971389771
6 0.680000066757202
5 0.674999952316284
4 0.700000047683716
3 0.660000085830688
2 0.924999952316284
1 0.690000057220459
};
\addplot [semithick, cornflowerblue109146255]
table {%
18 0.680000066757202
17 0.575000047683716
16 0.779999971389771
15 0.475000023841858
14 0.700000047683716
13 0.825000047683716
12 0.910000085830688
11 0.934999942779541
10 0.889999985694885
9 0.409999966621399
8 0.774999976158142
7 0.514999985694885
6 0.610000014305115
5 0.634999990463257
4 0.639999985694885
3 0.674999952316284
2 0.875
1 0.620000004768372
};
\addplot [semithick, mediumslateblue146109255]
table {%
18 0.615000009536743
17 0.504999995231628
16 0.759999990463257
15 0.450000047683716
14 0.674999952316284
13 0.815000057220459
12 0.914999961853027
11 0.879999995231628
10 0.920000076293945
9 0.389999985694885
8 0.809999942779541
7 0.559999942779541
6 0.710000038146973
5 0.639999985694885
4 0.684999942779541
3 0.690000057220459
2 0.879999995231628
1 0.694999933242798
};
\addplot [semithick, mediumorchid18273255]
table {%
18 0.5
17 0.5
16 0.644999980926514
15 0.485000014305115
14 0.585000038146973
13 0.795000076293945
12 0.839999914169312
10 0.920000076293945
9 0.389999985694885
8 0.759999990463257
7 0.549999952316284
6 0.660000085830688
5 0.585000038146973
3 0.664999961853027
2 0.875
1 0.589999914169312
};
\addplot [semithick, magenta21936255]
table {%
18 0.450000047683716
17 0.394999980926514
16 0.625
15 0.414999961853027
14 0.545000076293945
13 0.725000023841858
12 0.819999933242798
11 0.835000038146973
10 0.910000085830688
9 0.399999976158142
8 0.694999933242798
7 0.600000023841858
6 0.700000047683716
5 0.710000038146973
4 0.649999976158142
3 0.605000019073486
2 0.924999952316284
1 0.585000038146973
};
\addplot [semithick, magenta]
table {%
18 0.0149999856948853
17 0.0399999618530273
16 0.0349999666213989
15 0.0449999570846558
14 0.159999966621399
13 0.184999942779541
12 0.299999952316284
11 0.559999942779541
10 0.615000009536743
9 0.269999980926514
8 0.610000014305115
7 0.485000014305115
6 0.620000004768372
5 0.549999952316284
4 0.579999923706055
3 0.625
2 0.855000019073486
1 0.600000023841858
};
\legend{ , , , , , ,, 0,0.2, 0.5, 1, 2,5,10,100}
\end{axis}

\end{tikzpicture}
\begin{tikzpicture}[scale=0.8]

\definecolor{cornflowerblue109146255}{RGB}{109,146,255}
\definecolor{cornflowerblue73182255}{RGB}{73,182,255}
\definecolor{cyan}{RGB}{0,255,255}
\definecolor{darkgray176}{RGB}{176,176,176}
\definecolor{deepskyblue36219255}{RGB}{36,219,255}
\definecolor{magenta}{RGB}{255,0,255}
\definecolor{magenta21936255}{RGB}{219,36,255}
\definecolor{mediumorchid18273255}{RGB}{182,73,255}
\definecolor{mediumslateblue146109255}{RGB}{146,109,255}

\begin{axis}[
tick align=center,
tick pos=both,
title={HeH$^+$ memory error scaled},
x grid style={darkgray176},
xlabel={Readout qubits},
xmin=0, xmax=20,
xtick style={color=black},
xtick={0,5,10,15,20},
xticklabels={
  \(\displaystyle {0}\),
  \(\displaystyle {5}\),
  \(\displaystyle {10}\),
  \(\displaystyle {15}\),
  \(\displaystyle {20}\)
},
y grid style={darkgray176},
ylabel={P\(\displaystyle _{max}\)},
ymin=-0.05, ymax=1.05,
ytick style={color=black},
ytick={-0.2,0,0.2,0.4,0.6,0.8,1,1.2},
yticklabels={
  \(\displaystyle {\ensuremath{-}0.2}\),
  \(\displaystyle {0.0}\),
  \(\displaystyle {0.2}\),
  \(\displaystyle {0.4}\),
  \(\displaystyle {0.6}\),
  \(\displaystyle {0.8}\),
  \(\displaystyle {1.0}\),
  \(\displaystyle {1.2}\)
}
]
\addplot [draw=deepskyblue36219255, fill=deepskyblue36219255, mark=x,mark size=4pt, only marks]
table{%
x  y
18 0.33
17 0.545
nan nan
15 0.55
14 0.415
13 0.52
12 0.435
11 0.525
10 0.745
9 0.45
8 0.7
7 0.795
6 0.84
5 0.375
4 0.77
3 0.86
2 0.935
};
\addplot [draw=cornflowerblue73182255, fill=cornflowerblue73182255, mark=x,mark size=4pt, only marks]
table{%
x  y
18 0.195
17 0.335
16 0.235
15 0.425
14 0.36
13 0.455
12 0.5
11 0.505
10 0.715
9 0.41
8 0.705
7 0.795
6 0.87
5 0.415
4 0.73
3 0.91
2 0.92
1 0.6
};
\addplot [draw=cornflowerblue109146255, fill=cornflowerblue109146255, mark=x,mark size=4pt, only marks]
table{%
x  y
18 0.0649999999999999
17 0.155
nan nan
15 0.385
14 0.295
13 0.44
12 0.375
11 0.525
10 0.65
9 0.36
8 0.67
7 0.84
6 0.84
5 0.415
4 0.75
3 0.895
2 0.965
1 0.575
};
\addplot [draw=mediumslateblue146109255, fill=mediumslateblue146109255, mark=x,mark size=4pt, only marks]
table{%
x  y
17 0.0700000000000001
16 0.0649999999999999
15 0.245
14 0.23
13 0.275
12 0.295
11 0.4
10 0.575
9 0.345
8 0.635
7 0.745
6 0.775
5 0.36
4 0.755
3 0.875
2 0.92
1 0.555
};
\addplot [draw=mediumorchid18273255, fill=mediumorchid18273255, mark=x,mark size=4pt, only marks]
table{%
x  y
14 0.0549999999999999
nan nan
12 0.22
11 0.27
10 0.46
9 0.38
8 0.52
7 0.725
6 0.78
5 0.375
4 0.74
3 0.865
2 0.925
1 0.52
};
\addplot [draw=magenta21936255, fill=magenta21936255, mark=x,mark size=4pt, only marks]
table{%
x  y
11 0.185
10 0.32
9 0.215
8 0.505
7 0.66
6 0.73
5 0.37
4 0.71
3 0.885
2 0.93
1 0.575
};
\addplot [draw=magenta, fill=magenta, mark=x,mark size=4pt, only marks]
table{%
x  y
4 0.47
3 0.57
2 0.78
1 0.51
};
\addplot [semithick, cyan]
table {%
18 0.535000085830688
17 0.629999995231628
16 0.345000028610229
15 0.585000038146973
14 0.460000038146973
13 0.504999995231628
12 0.470000028610229
11 0.559999942779541
10 0.714999914169312
9 0.490000009536743
8 0.75
7 0.850000023841858
6 0.839999914169312
5 0.470000028610229
4 0.789999961853027
3 0.875
2 0.934999942779541
1 0.539999961853027
};
\addplot [semithick, deepskyblue36219255]
table {%
18 0.330000042915344
17 0.545000076293945
16 0.215000033378601
15 0.549999952316284
14 0.414999961853027
13 0.519999980926514
12 0.434999942779541
11 0.524999976158142
10 0.745000004768372
9 0.450000047683716
8 0.700000047683716
7 0.795000076293945
6 0.839999914169312
5 0.375
4 0.769999980926514
3 0.860000014305115
2 0.934999942779541
1 0.5
};
\addplot [semithick, cornflowerblue73182255]
table {%
18 0.194999933242798
17 0.335000038146973
16 0.235000014305115
15 0.424999952316284
14 0.360000014305115
13 0.455000042915344
12 0.5
11 0.504999995231628
10 0.714999914169312
9 0.409999966621399
8 0.704999923706055
7 0.795000076293945
6 0.870000004768372
5 0.414999961853027
4 0.730000019073486
3 0.910000085830688
2 0.920000076293945
1 0.600000023841858
};
\addplot [semithick, cornflowerblue109146255]
table {%
18 0.065000057220459
17 0.154999971389771
16 0.120000004768372
15 0.384999990463257
14 0.294999957084656
13 0.440000057220459
12 0.375
11 0.524999976158142
10 0.649999976158142
9 0.360000014305115
8 0.670000076293945
7 0.839999914169312
6 0.839999914169312
5 0.414999961853027
4 0.75
3 0.894999980926514
2 0.964999914169312
1 0.575000047683716
};
\addplot [semithick, mediumslateblue146109255]
table {%
18 0.0249999761581421
17 0.0700000524520874
16 0.065000057220459
15 0.245000004768372
14 0.230000019073486
13 0.274999976158142
12 0.294999957084656
11 0.399999976158142
10 0.575000047683716
9 0.345000028610229
8 0.634999990463257
7 0.745000004768372
6 0.774999976158142
5 0.360000014305115
4 0.754999995231628
3 0.875
2 0.920000076293945
1 0.555000066757202
};
\addplot [semithick, mediumorchid18273255]
table {%
18 0.00999999046325684
17 0.0149999856948853
16 0.0399999618530273
15 0.0449999570846558
14 0.0549999475479126
13 0.0800000429153442
12 0.220000028610229
11 0.269999980926514
10 0.460000038146973
9 0.379999995231628
8 0.519999980926514
7 0.725000023841858
6 0.779999971389771
5 0.375
4 0.740000009536743
3 0.865000009536743
2 0.924999952316284
1 0.519999980926514
};
\addplot [semithick, magenta21936255]
table {%
18 0.00999999046325684
17 0.0149999856948853
16 0.0149999856948853
14 0.0349999666213989
12 0.065000057220459
11 0.184999942779541
10 0.319999933242798
9 0.215000033378601
8 0.504999995231628
7 0.660000085830688
6 0.730000019073486
5 0.370000004768372
4 0.710000038146973
3 0.884999990463257
2 0.930000066757202
1 0.575000047683716
};
\addplot [semithick, magenta]
table {%
18 0.00499999523162842
16 0.00499999523162842
14 0.0149999856948853
13 0.00999999046325684
12 0.0149999856948853
11 0.00999999046325684
9 0.0299999713897705
8 0.065000057220459
7 0.0750000476837158
6 0.139999985694885
5 0.144999980926514
4 0.470000028610229
3 0.569999933242798
2 0.779999971389771
1 0.509999990463257
};
\end{axis}

\end{tikzpicture}
    \caption{Probability to read the most likely phase for between 1 and 18 readout ancillae with the H2 ion-trap emulator memory (dephasing) error scaled by between 0 (blue, top) and 100 (pink, bottom). `$\times$' marks denote results where the highest probability phase of the noisy result matches the highest probability phase of the memory error-free result for 200 shots. }
    \label{memory_scaled_pmax}
\end{figure}
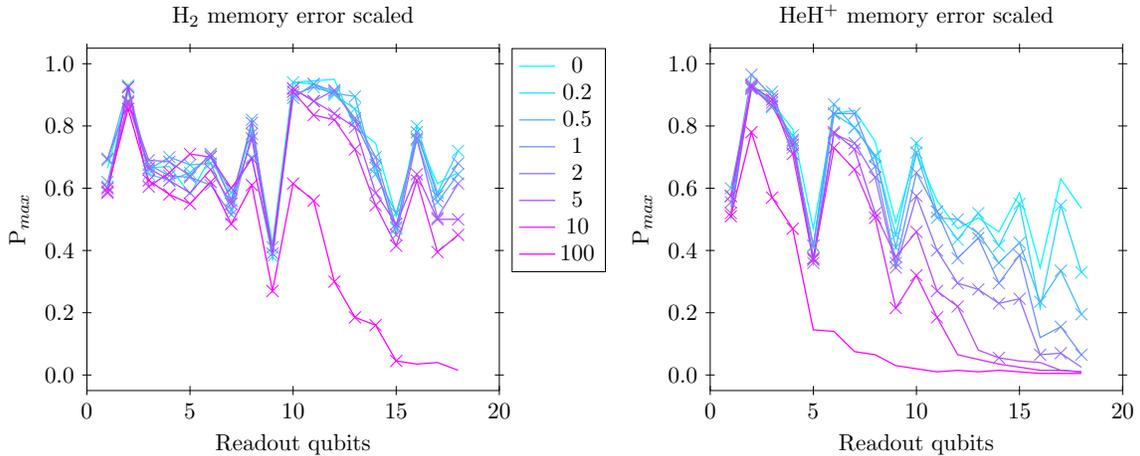

\begin{figure}[h]
    \centering
\begin{tikzpicture}[scale=0.8]

\definecolor{cornflowerblue109146255}{RGB}{109,146,255}
\definecolor{cornflowerblue73182255}{RGB}{73,182,255}
\definecolor{cyan}{RGB}{0,255,255}
\definecolor{darkgray176}{RGB}{176,176,176}
\definecolor{deepskyblue36219255}{RGB}{36,219,255}
\definecolor{magenta}{RGB}{255,0,255}
\definecolor{magenta21936255}{RGB}{219,36,255}
\definecolor{mediumorchid18273255}{RGB}{182,73,255}
\definecolor{mediumslateblue146109255}{RGB}{146,109,255}

\begin{axis}[
tick align=center,
tick pos=both,
legend pos = outer north east,
title={H$_2$ memory error scaled},
x grid style={darkgray176},
xlabel={Readout qubits},
xmin=0, xmax=20,
xtick style={color=black},
xtick={0,5,10,15,20},
xticklabels={
  \(\displaystyle {0}\),
  \(\displaystyle {5}\),
  \(\displaystyle {10}\),
  \(\displaystyle {15}\),
  \(\displaystyle {20}\)
},
y grid style={darkgray176},
ylabel={P\(\displaystyle _{sum\, of\, 3}\)},
ymin=-0.01325, ymax=1.04825,
ytick style={color=black},
ytick={-0.2,0,0.2,0.4,0.6,0.8,1,1.2},
yticklabels={
  \(\displaystyle {\ensuremath{-}0.2}\),
  \(\displaystyle {0.0}\),
  \(\displaystyle {0.2}\),
  \(\displaystyle {0.4}\),
  \(\displaystyle {0.6}\),
  \(\displaystyle {0.8}\),
  \(\displaystyle {1.0}\),
  \(\displaystyle {1.2}\)
}
]
\addplot [draw=deepskyblue36219255, fill=deepskyblue36219255, mark=x,mark size=4pt, only marks]
table{%
x  y
18 0.795
17 0.82
16 0.885
15 0.805
14 0.85
13 0.895
12 0.93
11 0.96
10 0.96
9 0.81
8 0.855
7 0.84
6 0.87
5 0.845
4 0.875
3 0.895
2 1
1 1
};
\addplot [draw=cornflowerblue73182255, fill=cornflowerblue73182255, mark=x,mark size=4pt, only marks]
table{%
x  y
18 0.77
17 0.755
16 0.875
15 0.795
14 0.805
13 0.945
12 0.94
11 0.97
10 0.94
nan nan
8 0.905
7 0.845
6 0.86
5 0.875
4 0.895
3 0.91
2 0.99
1 1
};
\addplot [draw=cornflowerblue109146255, fill=cornflowerblue109146255, mark=x,mark size=4pt, only marks]
table{%
x  y
18 0.825
17 0.8
16 0.855
15 0.77
14 0.825
13 0.875
12 0.94
11 0.955
10 0.925
9 0.785
8 0.875
7 0.825
6 0.84
5 0.9
4 0.895
3 0.905
2 0.98
1 1
};
\addplot [draw=mediumslateblue146109255, fill=mediumslateblue146109255, mark=x,mark size=4pt, only marks]
table{%
x  y
18 0.705
17 0.745
16 0.835
15 0.72
14 0.815
13 0.875
12 0.945
11 0.925
10 0.945
9 0.795
8 0.91
7 0.85
6 0.89
5 0.875
4 0.895
3 0.895
2 0.985
1 1
};
\addplot [draw=mediumorchid18273255, fill=mediumorchid18273255, mark=x,mark size=4pt, only marks]
table{%
x  y
18 0.6
17 0.655
16 0.69
15 0.72
14 0.76
13 0.865
12 0.88
11 0.92
10 0.95
nan nan
8 0.92
7 0.805
6 0.855
5 0.86
4 0.855
3 0.87
2 0.965
1 1
};
\addplot [draw=magenta21936255, fill=magenta21936255, mark=x,mark size=4pt, only marks]
table{%
x  y
18 0.51
17 0.54
16 0.675
15 0.62
14 0.73
13 0.795
12 0.87
11 0.88
10 0.945
nan nan
8 0.84
7 0.82
6 0.88
5 0.87
4 0.865
3 0.885
2 0.985
1 1
};
\addplot [draw=magenta, fill=magenta, mark=x,mark size=4pt, only marks]
table{%
x  y
15 0.115
14 0.245
13 0.34
12 0.455
11 0.67
10 0.72
9 0.575
8 0.69
7 0.7
6 0.8
5 0.755
4 0.805
3 0.885
2 0.99
1 1
};
\addplot [semithick, cyan]
table {%
18 0.799999952316284
17 0.829999923706055
16 0.870000004768372
15 0.759999990463257
14 0.875
13 0.875
12 0.960000038146973
11 0.980000019073486
10 0.970000028610229
9 0.789999961853027
8 0.865000009536743
7 0.829999923706055
6 0.889999985694885
5 0.819999933242798
4 0.875
3 0.884999990463257
2 0.975000023841858
1 1
};
\addplot [semithick, deepskyblue36219255]
table {%
18 0.795000076293945
17 0.819999933242798
16 0.884999990463257
15 0.805000066757202
13 0.894999980926514
12 0.930000066757202
11 0.960000038146973
10 0.960000038146973
9 0.809999942779541
8 0.855000019073486
7 0.839999914169312
6 0.870000004768372
5 0.845000028610229
4 0.875
3 0.894999980926514
2 1
1 1
};
\addplot [semithick, cornflowerblue73182255]
table {%
18 0.769999980926514
17 0.754999995231628
16 0.875
15 0.795000076293945
14 0.805000066757202
13 0.944999933242798
12 0.940000057220459
11 0.970000028610229
10 0.940000057220459
9 0.774999976158142
8 0.904999971389771
7 0.845000028610229
5 0.875
4 0.894999980926514
3 0.910000085830688
2 0.990000009536743
1 1
};
\addplot [semithick, cornflowerblue109146255]
table {%
18 0.825000047683716
17 0.799999952316284
16 0.855000019073486
15 0.769999980926514
14 0.825000047683716
13 0.875
12 0.940000057220459
11 0.954999923706055
10 0.924999952316284
9 0.785000085830688
8 0.875
7 0.825000047683716
6 0.839999914169312
5 0.899999976158142
4 0.894999980926514
3 0.904999971389771
2 0.980000019073486
1 1
};
\addplot [semithick, mediumslateblue146109255]
table {%
18 0.704999923706055
17 0.745000004768372
16 0.835000038146973
15 0.720000028610229
14 0.815000057220459
13 0.875
12 0.944999933242798
11 0.924999952316284
10 0.944999933242798
9 0.795000076293945
8 0.910000085830688
7 0.850000023841858
6 0.889999985694885
5 0.875
4 0.894999980926514
3 0.894999980926514
2 0.985000014305115
1 1
};
\addplot [semithick, mediumorchid18273255]
table {%
18 0.600000023841858
17 0.654999971389771
16 0.690000057220459
15 0.720000028610229
14 0.759999990463257
13 0.865000009536743
12 0.879999995231628
11 0.920000076293945
10 0.950000047683716
9 0.799999952316284
8 0.920000076293945
7 0.805000066757202
6 0.855000019073486
5 0.860000014305115
4 0.855000019073486
3 0.870000004768372
2 0.964999914169312
1 1
};
\addplot [semithick, magenta21936255]
table {%
18 0.509999990463257
17 0.539999961853027
16 0.674999952316284
15 0.620000004768372
14 0.730000019073486
13 0.795000076293945
12 0.870000004768372
11 0.879999995231628
10 0.944999933242798
9 0.769999980926514
8 0.839999914169312
7 0.819999933242798
6 0.879999995231628
5 0.870000004768372
4 0.865000009536743
3 0.884999990463257
2 0.985000014305115
1 1
};
\addplot [semithick, magenta]
table {%
18 0.0349999666213989
17 0.0800000429153442
16 0.0900000333786011
15 0.115000009536743
14 0.245000004768372
13 0.340000033378601
12 0.455000042915344
11 0.670000076293945
10 0.720000028610229
9 0.575000047683716
8 0.690000057220459
7 0.700000047683716
6 0.799999952316284
5 0.754999995231628
4 0.805000066757202
3 0.884999990463257
2 0.990000009536743
1 1
};
\legend{ , , , , , ,, 0,0.2, 0.5, 1, 2,5,10,100}
\end{axis}

\end{tikzpicture}
\begin{tikzpicture}[scale=0.8]

\definecolor{cornflowerblue109146255}{RGB}{109,146,255}
\definecolor{cornflowerblue73182255}{RGB}{73,182,255}
\definecolor{cyan}{RGB}{0,255,255}
\definecolor{darkgray176}{RGB}{176,176,176}
\definecolor{deepskyblue36219255}{RGB}{36,219,255}
\definecolor{magenta}{RGB}{255,0,255}
\definecolor{magenta21936255}{RGB}{219,36,255}
\definecolor{mediumorchid18273255}{RGB}{182,73,255}
\definecolor{mediumslateblue146109255}{RGB}{146,109,255}

\begin{axis}[
tick align=center,
tick pos=both,
title={HeH$^+$ memory error scaled},
x grid style={darkgray176},
xlabel={Readout qubits},
xmin=0, xmax=20,
xtick style={color=black},
xtick={0,5,10,15,20},
xticklabels={
  \(\displaystyle {0}\),
  \(\displaystyle {5}\),
  \(\displaystyle {10}\),
  \(\displaystyle {15}\),
  \(\displaystyle {20}\)
},
y grid style={darkgray176},
ylabel={P\(\displaystyle _{sum\, of\, 3}\)},
ymin=-0.03425, ymax=1.04925,
ytick style={color=black},
ytick={-0.2,0,0.2,0.4,0.6,0.8,1,1.2},
yticklabels={
  \(\displaystyle {\ensuremath{-}0.2}\),
  \(\displaystyle {0.0}\),
  \(\displaystyle {0.2}\),
  \(\displaystyle {0.4}\),
  \(\displaystyle {0.6}\),
  \(\displaystyle {0.8}\),
  \(\displaystyle {1.0}\),
  \(\displaystyle {1.2}\)
}
]
\addplot [draw=deepskyblue36219255, fill=deepskyblue36219255, mark=x,mark size=4pt, only marks]
table{%
x  y
18 0.39
17 0.575
nan nan
15 0.67
14 0.62
13 0.615
12 0.585
11 0.69
10 0.8
9 0.71
8 0.785
7 0.86
6 0.885
5 0.795
4 0.86
3 0.94
2 0.985
};
\addplot [draw=cornflowerblue73182255, fill=cornflowerblue73182255, mark=x,mark size=4pt, only marks]
table{%
x  y
18 0.255
17 0.405
16 0.405
15 0.52
14 0.55
13 0.58
12 0.645
11 0.66
10 0.77
9 0.66
8 0.785
7 0.835
6 0.895
5 0.81
4 0.87
3 0.96
2 0.985
1 1
};
\addplot [draw=cornflowerblue109146255, fill=cornflowerblue109146255, mark=x,mark size=4pt, only marks]
table{%
x  y
18 0.13
17 0.245
nan nan
15 0.455
14 0.415
13 0.57
12 0.52
11 0.72
10 0.715
9 0.67
8 0.765
7 0.89
6 0.9
5 0.8
4 0.895
3 0.95
2 0.995
1 1
};
\addplot [draw=mediumslateblue146109255, fill=mediumslateblue146109255, mark=x,mark size=4pt, only marks]
table{%
x  y
17 0.165
16 0.11
15 0.36
14 0.34
13 0.37
12 0.435
11 0.565
10 0.64
9 0.605
8 0.765
7 0.815
6 0.835
5 0.75
4 0.875
3 0.94
2 0.975
1 1
};
\addplot [draw=mediumorchid18273255, fill=mediumorchid18273255, mark=x,mark size=4pt, only marks]
table{%
x  y
14 0.14
nan nan
12 0.355
11 0.395
10 0.545
9 0.605
8 0.63
7 0.78
6 0.86
5 0.755
4 0.865
3 0.945
2 0.985
1 1
};
\addplot [draw=magenta21936255, fill=magenta21936255, mark=x,mark size=4pt, only marks]
table{%
x  y
11 0.31
10 0.47
9 0.425
8 0.645
7 0.75
6 0.795
5 0.715
4 0.83
3 0.96
2 0.98
1 1
};
\addplot [draw=magenta, fill=magenta, mark=x,mark size=4pt, only marks]
table{%
x  y
4 0.7
3 0.78
2 0.95
1 1
};
\addplot [semithick, cyan]
table {%
18 0.615000009536743
17 0.700000047683716
16 0.634999990463257
15 0.684999942779541
14 0.690000057220459
12 0.639999985694885
11 0.735000014305115
10 0.785000085830688
9 0.774999976158142
8 0.809999942779541
7 0.910000085830688
6 0.894999980926514
5 0.829999923706055
4 0.894999980926514
2 0.985000014305115
1 1
};
\addplot [semithick, deepskyblue36219255]
table {%
18 0.389999985694885
17 0.575000047683716
16 0.485000014305115
15 0.670000076293945
14 0.620000004768372
13 0.615000009536743
12 0.585000038146973
11 0.690000057220459
10 0.799999952316284
9 0.710000038146973
7 0.860000014305115
6 0.884999990463257
5 0.795000076293945
4 0.860000014305115
3 0.940000057220459
2 0.985000014305115
1 1
};
\addplot [semithick, cornflowerblue73182255]
table {%
18 0.254999995231628
17 0.404999971389771
16 0.404999971389771
15 0.519999980926514
13 0.579999923706055
12 0.644999980926514
11 0.660000085830688
10 0.769999980926514
9 0.660000085830688
8 0.785000085830688
7 0.835000038146973
6 0.894999980926514
5 0.809999942779541
4 0.870000004768372
3 0.960000038146973
2 0.985000014305115
1 1
};
\addplot [semithick, cornflowerblue109146255]
table {%
18 0.129999995231628
17 0.245000004768372
16 0.269999980926514
15 0.455000042915344
14 0.414999961853027
13 0.569999933242798
12 0.519999980926514
11 0.720000028610229
10 0.714999914169312
9 0.670000076293945
8 0.764999985694885
7 0.889999985694885
6 0.899999976158142
5 0.799999952316284
4 0.894999980926514
3 0.950000047683716
2 0.995000004768372
1 1
};
\addplot [semithick, mediumslateblue146109255]
table {%
18 0.0700000524520874
17 0.164999961853027
16 0.110000014305115
15 0.360000014305115
14 0.340000033378601
13 0.370000004768372
12 0.434999942779541
11 0.565000057220459
10 0.639999985694885
9 0.605000019073486
8 0.764999985694885
7 0.815000057220459
6 0.835000038146973
5 0.75
4 0.875
3 0.940000057220459
2 0.975000023841858
1 1
};
\addplot [semithick, mediumorchid18273255]
table {%
18 0.0299999713897705
17 0.0449999570846558
16 0.0800000429153442
15 0.129999995231628
14 0.139999985694885
13 0.205000042915344
12 0.355000019073486
11 0.394999980926514
10 0.545000076293945
9 0.605000019073486
8 0.629999995231628
7 0.779999971389771
6 0.860000014305115
5 0.754999995231628
4 0.865000009536743
3 0.944999933242798
2 0.985000014305115
1 1
};
\addplot [semithick, magenta21936255]
table {%
18 0.0299999713897705
16 0.0399999618530273
15 0.0750000476837158
14 0.0900000333786011
13 0.134999990463257
12 0.159999966621399
11 0.309999942779541
10 0.470000028610229
9 0.424999952316284
8 0.644999980926514
7 0.75
6 0.795000076293945
5 0.714999914169312
4 0.829999923706055
3 0.960000038146973
1 1
};
\addplot [semithick, magenta]
table {%
18 0.0149999856948853
16 0.0149999856948853
15 0.0199999809265137
14 0.0349999666213989
13 0.0299999713897705
12 0.0349999666213989
11 0.0299999713897705
10 0.059999942779541
9 0.0750000476837158
8 0.139999985694885
7 0.210000038146973
6 0.375
5 0.355000019073486
4 0.700000047683716
3 0.779999971389771
2 0.950000047683716
1 1
};
\end{axis}

\end{tikzpicture}
    \caption{Probability to read the three most likely phases for between 1 and 18 readout ancillae with the H2 ion-trap emulator memory (dephasing) error scaled by between 0 (blue, top) and 100 (pink, bottom). `$\times$' marks denote results where the highest probability phase of the noisy result matches the highest probability phase of the memory error-free result for 200 shots.  }
    \label{memory_scaled_p3max}
\end{figure}
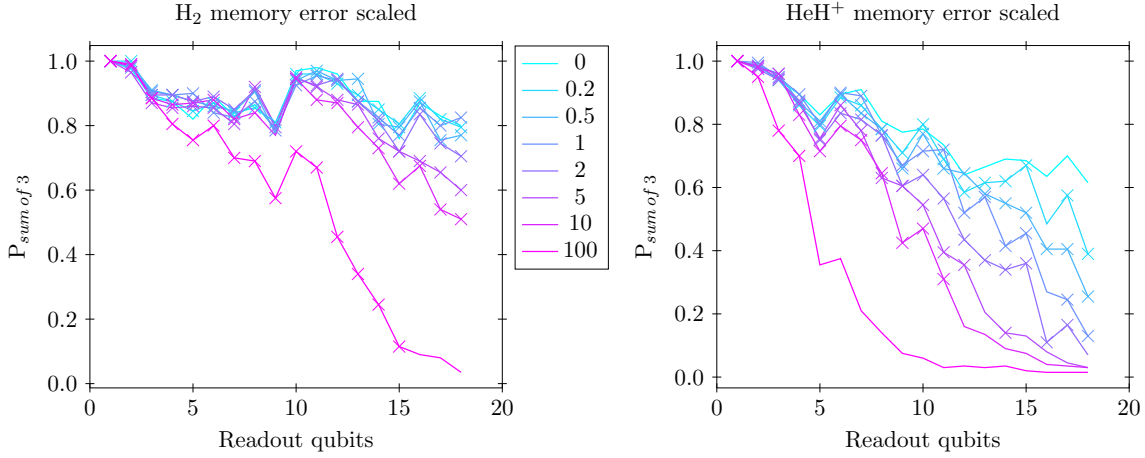

\newpage
\subsection{Two-qubit gate error scaling}

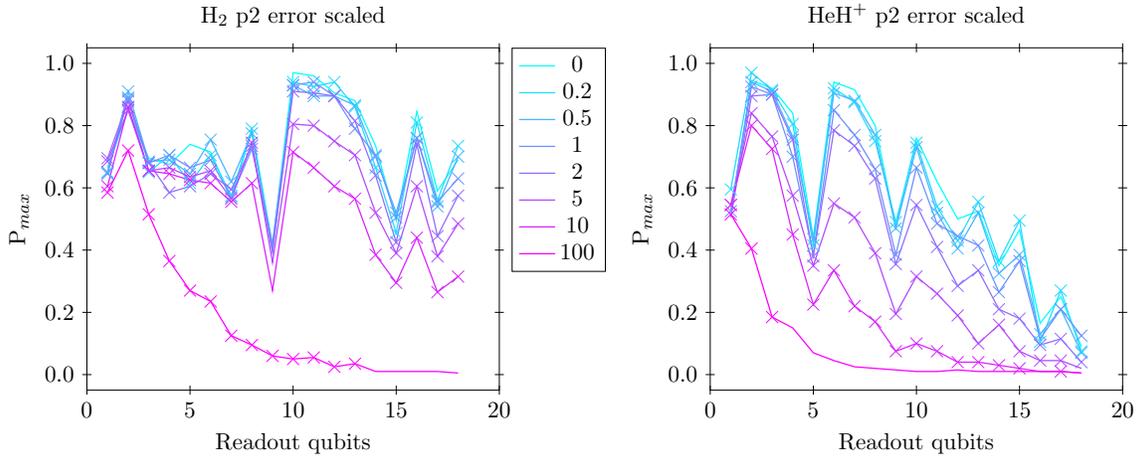
\begin{figure}[h]
    \centering
\begin{tikzpicture}[scale=0.8]

\definecolor{cornflowerblue109146255}{RGB}{109,146,255}
\definecolor{cornflowerblue73182255}{RGB}{73,182,255}
\definecolor{cyan}{RGB}{0,255,255}
\definecolor{darkgray176}{RGB}{176,176,176}
\definecolor{deepskyblue36219255}{RGB}{36,219,255}
\definecolor{magenta}{RGB}{255,0,255}
\definecolor{magenta21936255}{RGB}{219,36,255}
\definecolor{mediumorchid18273255}{RGB}{182,73,255}
\definecolor{mediumslateblue146109255}{RGB}{146,109,255}

\begin{axis}[
tick align=center,
tick pos=both,
legend pos = outer north east,
title={H$_2$ p2 error scaled},
x grid style={darkgray176},
xlabel={Readout qubits},
xmin=0, xmax=20,
xtick style={color=black},
xtick={0,5,10,15,20},
xticklabels={
  \(\displaystyle {0}\),
  \(\displaystyle {5}\),
  \(\displaystyle {10}\),
  \(\displaystyle {15}\),
  \(\displaystyle {20}\)
},
y grid style={darkgray176},
ylabel={P\(\displaystyle _{max}\)},
ymin=-0.05, ymax=1.05,
ytick style={color=black},
ytick={-0.2,0,0.2,0.4,0.6,0.8,1,1.2},
yticklabels={
  \(\displaystyle {\ensuremath{-}0.2}\),
  \(\displaystyle {0.0}\),
  \(\displaystyle {0.2}\),
  \(\displaystyle {0.4}\),
  \(\displaystyle {0.6}\),
  \(\displaystyle {0.8}\),
  \(\displaystyle {1.0}\),
  \(\displaystyle {1.2}\)
}
]
\addplot [draw=deepskyblue36219255, fill=deepskyblue36219255, mark=x,mark size=4pt, only marks]
table{%
x  y
18 0.735
17 0.56
16 0.81
15 0.455
14 0.665
13 0.86
12 0.94
11 0.925
10 0.94
nan nan
8 0.79
7 0.575
6 0.705
5 0.615
4 0.705
3 0.65
2 0.885
1 0.65
};
\addplot [draw=cornflowerblue73182255, fill=cornflowerblue73182255, mark=x,mark size=4pt, only marks]
table{%
x  y
18 0.7
17 0.54
16 0.76
15 0.52
14 0.7
13 0.865
12 0.895
11 0.895
10 0.93
nan nan
8 0.775
7 0.62
6 0.755
5 0.645
4 0.69
3 0.685
2 0.91
1 0.65
};
\addplot [draw=cornflowerblue109146255, fill=cornflowerblue109146255, mark=x,mark size=4pt, only marks]
table{%
x  y
18 0.63
17 0.555
16 0.76
15 0.505
14 0.705
13 0.79
12 0.895
11 0.94
10 0.93
nan nan
8 0.725
7 0.57
6 0.69
5 0.665
4 0.705
3 0.68
2 0.89
1 0.645
};
\addplot [draw=mediumslateblue146109255, fill=mediumslateblue146109255, mark=x,mark size=4pt, only marks]
table{%
x  y
18 0.575
17 0.445
16 0.74
15 0.425
14 0.64
13 0.815
12 0.895
11 0.905
10 0.91
nan nan
8 0.735
7 0.595
6 0.645
5 0.605
4 0.585
3 0.68
2 0.88
1 0.695
};
\addplot [draw=mediumorchid18273255, fill=mediumorchid18273255, mark=x,mark size=4pt, only marks]
table{%
x  y
18 0.485
17 0.38
16 0.605
15 0.39
14 0.52
13 0.705
12 0.75
11 0.8
10 0.805
nan nan
8 0.745
7 0.565
6 0.655
5 0.635
4 0.665
3 0.655
2 0.85
1 0.685
};
\addplot [draw=magenta21936255, fill=magenta21936255, mark=x,mark size=4pt, only marks]
table{%
x  y
18 0.315
17 0.265
16 0.44
15 0.295
14 0.385
13 0.565
12 0.605
11 0.665
10 0.715
nan nan
8 0.615
7 0.555
6 0.615
5 0.625
4 0.645
3 0.655
2 0.86
1 0.605
};
\addplot [draw=magenta, fill=magenta, mark=x,mark size=4pt, only marks]
table{%
x  y
13 0.0349999999999999
12 0.0249999999999999
11 0.0549999999999999
10 0.05
9 0.0600000000000001
8 0.095
7 0.125
6 0.235
5 0.27
4 0.365
3 0.515
2 0.72
1 0.585
};
\addplot [semithick, cyan]
table {%
18 0.690000057220459
17 0.589999914169312
16 0.845000028610229
15 0.495000004768372
14 0.735000014305115
13 0.879999995231628
12 0.904999971389771
11 0.960000038146973
10 0.970000028610229
9 0.414999961853027
8 0.785000085830688
7 0.565000057220459
6 0.714999914169312
5 0.740000009536743
4 0.674999952316284
3 0.694999933242798
2 0.850000023841858
1 0.664999961853027
};
\addplot [semithick, deepskyblue36219255]
table {%
18 0.735000014305115
17 0.559999942779541
16 0.809999942779541
15 0.455000042915344
14 0.664999961853027
13 0.860000014305115
12 0.940000057220459
11 0.924999952316284
10 0.940000057220459
9 0.409999966621399
8 0.789999961853027
7 0.575000047683716
6 0.704999923706055
5 0.615000009536743
4 0.704999923706055
3 0.649999976158142
2 0.884999990463257
1 0.649999976158142
};
\addplot [semithick, cornflowerblue73182255]
table {%
18 0.700000047683716
17 0.539999961853027
16 0.759999990463257
15 0.519999980926514
14 0.700000047683716
13 0.865000009536743
12 0.894999980926514
11 0.894999980926514
10 0.930000066757202
9 0.394999980926514
8 0.774999976158142
7 0.620000004768372
6 0.754999995231628
5 0.644999980926514
4 0.690000057220459
3 0.684999942779541
2 0.910000085830688
1 0.649999976158142
};
\addplot [semithick, cornflowerblue109146255]
table {%
18 0.629999995231628
17 0.555000066757202
16 0.759999990463257
15 0.504999995231628
14 0.704999923706055
13 0.789999961853027
12 0.894999980926514
11 0.940000057220459
10 0.930000066757202
9 0.419999957084656
8 0.725000023841858
7 0.569999933242798
6 0.690000057220459
5 0.664999961853027
4 0.704999923706055
3 0.680000066757202
2 0.889999985694885
1 0.644999980926514
};
\addplot [semithick, mediumslateblue146109255]
table {%
18 0.575000047683716
17 0.444999933242798
16 0.740000009536743
15 0.424999952316284
14 0.639999985694885
13 0.815000057220459
12 0.894999980926514
11 0.904999971389771
10 0.910000085830688
9 0.394999980926514
8 0.735000014305115
7 0.595000028610229
6 0.644999980926514
5 0.605000019073486
4 0.585000038146973
3 0.680000066757202
2 0.879999995231628
1 0.694999933242798
};
\addplot [semithick, mediumorchid18273255]
table {%
18 0.485000014305115
17 0.379999995231628
16 0.605000019073486
15 0.389999985694885
14 0.519999980926514
13 0.704999923706055
12 0.75
11 0.799999952316284
10 0.805000066757202
9 0.360000014305115
8 0.745000004768372
7 0.565000057220459
6 0.654999971389771
5 0.634999990463257
4 0.664999961853027
3 0.654999971389771
2 0.850000023841858
1 0.684999942779541
};
\addplot [semithick, magenta21936255]
table {%
18 0.315000057220459
17 0.264999985694885
16 0.440000057220459
15 0.294999957084656
14 0.384999990463257
13 0.565000057220459
12 0.605000019073486
11 0.664999961853027
10 0.714999914169312
9 0.269999980926514
8 0.615000009536743
7 0.555000066757202
6 0.615000009536743
5 0.625
4 0.644999980926514
3 0.654999971389771
2 0.860000014305115
1 0.605000019073486
};
\addplot [semithick, magenta]
table {%
18 0.00499999523162842
17 0.00999999046325684
14 0.00999999046325684
13 0.0349999666213989
12 0.0249999761581421
11 0.0549999475479126
10 0.0499999523162842
9 0.059999942779541
8 0.0950000286102295
7 0.125
6 0.235000014305115
5 0.269999980926514
4 0.365000009536743
3 0.514999985694885
2 0.720000028610229
1 0.585000038146973
};
\legend{ , , , , , ,, 0,0.2, 0.5, 1, 2,5,10,100}
\end{axis}

\end{tikzpicture}
\begin{tikzpicture}[scale=0.8]

\definecolor{cornflowerblue109146255}{RGB}{109,146,255}
\definecolor{cornflowerblue73182255}{RGB}{73,182,255}
\definecolor{cyan}{RGB}{0,255,255}
\definecolor{darkgray176}{RGB}{176,176,176}
\definecolor{deepskyblue36219255}{RGB}{36,219,255}
\definecolor{magenta}{RGB}{255,0,255}
\definecolor{magenta21936255}{RGB}{219,36,255}
\definecolor{mediumorchid18273255}{RGB}{182,73,255}
\definecolor{mediumslateblue146109255}{RGB}{146,109,255}

\begin{axis}[
tick align=center,
tick pos=both,
title={HeH$^+$ p2 error scaled},
x grid style={darkgray176},
xlabel={Readout qubits},
xmin=0, xmax=20,
xtick style={color=black},
xtick={0,5,10,15,20},
xticklabels={
  \(\displaystyle {0}\),
  \(\displaystyle {5}\),
  \(\displaystyle {10}\),
  \(\displaystyle {15}\),
  \(\displaystyle {20}\)
},
y grid style={darkgray176},
ylabel={P\(\displaystyle _{max}\)},
ymin=-0.05, ymax=1.05,
ytick style={color=black},
ytick={-0.2,0,0.2,0.4,0.6,0.8,1,1.2},
yticklabels={
  \(\displaystyle {\ensuremath{-}0.2}\),
  \(\displaystyle {0.0}\),
  \(\displaystyle {0.2}\),
  \(\displaystyle {0.4}\),
  \(\displaystyle {0.6}\),
  \(\displaystyle {0.8}\),
  \(\displaystyle {1.0}\),
  \(\displaystyle {1.2}\)
}
]
\addplot [draw=deepskyblue36219255, fill=deepskyblue36219255, mark=x,mark size=4pt, only marks]
table{%
x  y
18 0.075
17 0.27
16 0.105
15 0.495
14 0.365
13 0.555
12 0.43
11 0.54
10 0.745
9 0.49
8 0.75
7 0.875
6 0.92
5 0.415
4 0.805
3 0.915
2 0.97
1 0.595
};
\addplot [draw=cornflowerblue73182255, fill=cornflowerblue73182255, mark=x,mark size=4pt, only marks]
table{%
x  y
18 0.0700000000000001
17 0.21
16 0.13
15 0.385
14 0.325
13 0.52
12 0.405
11 0.52
10 0.73
9 0.47
8 0.77
7 0.88
6 0.905
5 0.435
4 0.765
3 0.91
2 0.94
};
\addplot [draw=cornflowerblue109146255, fill=cornflowerblue109146255, mark=x,mark size=4pt, only marks]
table{%
x  y
18 0.125
17 0.21
nan nan
15 0.365
14 0.265
13 0.415
12 0.445
11 0.485
10 0.665
9 0.385
8 0.66
7 0.77
6 0.85
nan nan
4 0.7
3 0.9
2 0.925
1 0.525
};
\addplot [draw=mediumslateblue146109255, fill=mediumslateblue146109255, mark=x,mark size=4pt, only marks]
table{%
x  y
18 0.04
17 0.115
16 0.095
15 0.18
14 0.21
13 0.335
12 0.285
11 0.41
10 0.545
9 0.355
8 0.63
7 0.735
6 0.785
5 0.38
4 0.755
3 0.9
2 0.895
};
\addplot [draw=mediumorchid18273255, fill=mediumorchid18273255, mark=x,mark size=4pt, only marks]
table{%
x  y
17 0.0449999999999999
16 0.0449999999999999
15 0.075
14 0.16
13 0.1
12 0.19
11 0.26
10 0.315
9 0.195
8 0.39
7 0.505
6 0.55
5 0.35
4 0.575
3 0.765
2 0.84
1 0.545
};
\addplot [draw=magenta21936255, fill=magenta21936255, mark=x,mark size=4pt, only marks]
table{%
x  y
17 0.01
nan nan
15 0.02
14 0.03
13 0.04
12 0.04
11 0.075
10 0.1
9 0.075
8 0.17
7 0.22
6 0.335
5 0.225
4 0.45
3 0.725
2 0.8
1 0.545
};
\addplot [draw=magenta, fill=magenta, mark=x,mark size=4pt, only marks]
table{%
x  y
3 0.185
2 0.405
1 0.515
};
\addplot [semithick, cyan]
table {%
18 0.0900000333786011
17 0.25
16 0.164999961853027
15 0.465000033378601
14 0.350000023841858
13 0.524999976158142
12 0.5
11 0.620000004768372
10 0.759999990463257
9 0.475000023841858
8 0.799999952316284
7 0.914999961853027
6 0.940000057220459
5 0.440000057220459
4 0.839999914169312
3 0.920000076293945
2 0.944999933242798
1 0.509999990463257
};
\addplot [semithick, deepskyblue36219255]
table {%
18 0.0750000476837158
17 0.269999980926514
16 0.105000019073486
15 0.495000004768372
14 0.365000009536743
13 0.555000066757202
12 0.430000066757202
11 0.539999961853027
10 0.745000004768372
9 0.490000009536743
8 0.75
7 0.875
6 0.920000076293945
5 0.414999961853027
4 0.805000066757202
3 0.914999961853027
2 0.970000028610229
1 0.595000028610229
};
\addplot [semithick, cornflowerblue73182255]
table {%
18 0.0700000524520874
17 0.210000038146973
16 0.129999995231628
15 0.384999990463257
14 0.325000047683716
13 0.519999980926514
12 0.404999971389771
11 0.519999980926514
10 0.730000019073486
9 0.470000028610229
8 0.769999980926514
7 0.879999995231628
6 0.904999971389771
5 0.434999942779541
4 0.764999985694885
3 0.910000085830688
2 0.940000057220459
1 0.524999976158142
};
\addplot [semithick, cornflowerblue109146255]
table {%
18 0.125
17 0.210000038146973
16 0.115000009536743
15 0.365000009536743
14 0.264999985694885
13 0.414999961853027
12 0.444999933242798
11 0.485000014305115
10 0.664999961853027
9 0.384999990463257
8 0.660000085830688
7 0.769999980926514
6 0.850000023841858
5 0.384999990463257
4 0.700000047683716
3 0.899999976158142
2 0.924999952316284
1 0.524999976158142
};
\addplot [semithick, mediumslateblue146109255]
table {%
18 0.0399999618530273
17 0.115000009536743
16 0.0950000286102295
15 0.180000066757202
14 0.210000038146973
13 0.335000038146973
12 0.284999966621399
11 0.409999966621399
10 0.545000076293945
9 0.355000019073486
8 0.629999995231628
7 0.735000014305115
6 0.785000085830688
5 0.379999995231628
4 0.754999995231628
3 0.899999976158142
2 0.894999980926514
1 0.504999995231628
};
\addplot [semithick, mediumorchid18273255]
table {%
18 0.0199999809265137
17 0.0449999570846558
16 0.0449999570846558
15 0.0750000476837158
14 0.159999966621399
13 0.100000023841858
12 0.190000057220459
11 0.259999990463257
10 0.315000057220459
9 0.194999933242798
8 0.389999985694885
7 0.504999995231628
6 0.549999952316284
5 0.350000023841858
4 0.575000047683716
3 0.764999985694885
2 0.839999914169312
1 0.545000076293945
};
\addplot [semithick, magenta21936255]
table {%
18 0.00499999523162842
17 0.00999999046325684
16 0.00999999046325684
13 0.0399999618530273
12 0.0399999618530273
11 0.0750000476837158
10 0.100000023841858
9 0.0750000476837158
8 0.169999957084656
7 0.220000028610229
6 0.335000038146973
5 0.225000023841858
4 0.450000047683716
3 0.725000023841858
2 0.799999952316284
1 0.545000076293945
};
\addplot [semithick, magenta]
table {%
18 0.00499999523162842
17 0.00999999046325684
13 0.00999999046325684
12 0.0149999856948853
11 0.00999999046325684
10 0.00999999046325684
7 0.0249999761581421
6 0.0449999570846558
5 0.0700000524520874
4 0.149999976158142
3 0.184999942779541
2 0.404999971389771
1 0.514999985694885
};
\end{axis}

\end{tikzpicture}
    \caption{Probability to read the most likely phase for between 1 and 18 readout ancillae with the H2 ion-trap emulator two-qubit gate error rate (p2) scaled by between 0 (blue, top) and 100 (pink, bottom). `$\times$' marks denote results where the highest probability phase of the noisy result matches the highest probability phase of the memory error-free result for 200 shots.  }
    \label{p2_scaled_pmax}
\end{figure}

\begin{figure}[h]
    \centering
\begin{tikzpicture}[scale=0.8]

\definecolor{cornflowerblue109146255}{RGB}{109,146,255}
\definecolor{cornflowerblue73182255}{RGB}{73,182,255}
\definecolor{cyan}{RGB}{0,255,255}
\definecolor{darkgray176}{RGB}{176,176,176}
\definecolor{deepskyblue36219255}{RGB}{36,219,255}
\definecolor{magenta}{RGB}{255,0,255}
\definecolor{magenta21936255}{RGB}{219,36,255}
\definecolor{mediumorchid18273255}{RGB}{182,73,255}
\definecolor{mediumslateblue146109255}{RGB}{146,109,255}

\begin{axis}[
tick align=center,
tick pos=both,
legend pos = outer north east,
title={H$_2$ p2 error scaled},
x grid style={darkgray176},
xlabel={Readout qubits},
xmin=0, xmax=20,
xtick style={color=black},
xtick={0,5,10,15,20},
xticklabels={
  \(\displaystyle {0}\),
  \(\displaystyle {5}\),
  \(\displaystyle {10}\),
  \(\displaystyle {15}\),
  \(\displaystyle {20}\)
},
y grid style={darkgray176},
ylabel={P\(\displaystyle _{sum\, of\, 3}\)},
ymin=-0.05, ymax=1.05,
ytick style={color=black},
ytick={-0.2,0,0.2,0.4,0.6,0.8,1,1.2},
yticklabels={
  \(\displaystyle {\ensuremath{-}0.2}\),
  \(\displaystyle {0.0}\),
  \(\displaystyle {0.2}\),
  \(\displaystyle {0.4}\),
  \(\displaystyle {0.6}\),
  \(\displaystyle {0.8}\),
  \(\displaystyle {1.0}\),
  \(\displaystyle {1.2}\)
}
]
\addplot [draw=deepskyblue36219255, fill=deepskyblue36219255, mark=x,mark size=4pt, only marks]
table{%
x  y
18 0.865
17 0.81
16 0.875
15 0.795
14 0.87
13 0.915
12 0.965
11 0.955
10 0.96
nan nan
8 0.93
7 0.87
6 0.895
5 0.86
4 0.93
3 0.88
2 0.97
1 1
};
\addplot [draw=cornflowerblue73182255, fill=cornflowerblue73182255, mark=x,mark size=4pt, only marks]
table{%
x  y
18 0.785
17 0.79
16 0.83
15 0.81
14 0.865
13 0.905
12 0.93
11 0.93
10 0.965
nan nan
8 0.89
7 0.88
6 0.915
5 0.865
4 0.9
3 0.91
2 0.985
1 1
};
\addplot [draw=cornflowerblue109146255, fill=cornflowerblue109146255, mark=x,mark size=4pt, only marks]
table{%
x  y
18 0.775
17 0.765
16 0.815
15 0.795
14 0.855
13 0.85
12 0.945
11 0.97
10 0.965
nan nan
8 0.855
7 0.83
6 0.885
5 0.85
4 0.905
3 0.915
2 0.99
1 1
};
\addplot [draw=mediumslateblue146109255, fill=mediumslateblue146109255, mark=x,mark size=4pt, only marks]
table{%
x  y
18 0.72
17 0.67
16 0.82
15 0.725
14 0.785
13 0.885
12 0.925
11 0.93
10 0.945
nan nan
8 0.875
7 0.85
6 0.83
5 0.865
4 0.87
3 0.895
2 0.97
1 1
};
\addplot [draw=mediumorchid18273255, fill=mediumorchid18273255, mark=x,mark size=4pt, only marks]
table{%
x  y
18 0.595
17 0.555
16 0.66
15 0.61
14 0.67
13 0.765
12 0.8
11 0.835
10 0.84
nan nan
8 0.855
7 0.795
6 0.82
5 0.81
4 0.88
3 0.865
2 0.96
1 1
};
\addplot [draw=magenta21936255, fill=magenta21936255, mark=x,mark size=4pt, only marks]
table{%
x  y
18 0.395
17 0.42
16 0.495
15 0.52
14 0.49
13 0.605
12 0.66
11 0.715
10 0.76
nan nan
8 0.745
7 0.77
6 0.79
5 0.845
4 0.815
3 0.905
2 0.975
1 1
};
\addplot [draw=magenta, fill=magenta, mark=x,mark size=4pt, only marks]
table{%
x  y
13 0.0549999999999999
12 0.0549999999999999
11 0.11
10 0.1
9 0.115
8 0.2
7 0.215
6 0.36
5 0.445
4 0.63
3 0.75
2 0.925
1 1
};
\addplot [semithick, cyan]
table {%
18 0.829999923706055
17 0.799999952316284
16 0.920000076293945
15 0.855000019073486
14 0.870000004768372
13 0.930000066757202
12 0.934999942779541
11 0.985000014305115
10 0.990000009536743
9 0.839999914169312
8 0.899999976158142
7 0.855000019073486
6 0.914999961853027
4 0.875
3 0.894999980926514
2 0.960000038146973
1 1
};
\addplot [semithick, deepskyblue36219255]
table {%
18 0.865000009536743
17 0.809999942779541
16 0.875
15 0.795000076293945
14 0.870000004768372
13 0.914999961853027
12 0.964999914169312
11 0.954999923706055
10 0.960000038146973
9 0.829999923706055
8 0.930000066757202
7 0.870000004768372
6 0.894999980926514
5 0.860000014305115
4 0.930000066757202
3 0.879999995231628
2 0.970000028610229
1 1
};
\addplot [semithick, cornflowerblue73182255]
table {%
18 0.785000085830688
17 0.789999961853027
16 0.829999923706055
15 0.809999942779541
14 0.865000009536743
13 0.904999971389771
12 0.930000066757202
11 0.930000066757202
10 0.964999914169312
9 0.825000047683716
8 0.889999985694885
7 0.879999995231628
6 0.914999961853027
5 0.865000009536743
4 0.899999976158142
3 0.910000085830688
2 0.985000014305115
1 1
};
\addplot [semithick, cornflowerblue109146255]
table {%
18 0.774999976158142
17 0.764999985694885
16 0.815000057220459
15 0.795000076293945
14 0.855000019073486
13 0.850000023841858
12 0.944999933242798
11 0.970000028610229
10 0.964999914169312
9 0.819999933242798
8 0.855000019073486
7 0.829999923706055
6 0.884999990463257
5 0.850000023841858
4 0.904999971389771
3 0.914999961853027
2 0.990000009536743
1 1
};
\addplot [semithick, mediumslateblue146109255]
table {%
18 0.720000028610229
17 0.670000076293945
16 0.819999933242798
15 0.725000023841858
14 0.785000085830688
13 0.884999990463257
12 0.924999952316284
11 0.930000066757202
10 0.944999933242798
9 0.799999952316284
8 0.875
7 0.850000023841858
6 0.829999923706055
5 0.865000009536743
4 0.870000004768372
3 0.894999980926514
2 0.970000028610229
1 1
};
\addplot [semithick, mediumorchid18273255]
table {%
18 0.595000028610229
17 0.555000066757202
16 0.660000085830688
15 0.610000014305115
14 0.670000076293945
13 0.764999985694885
11 0.835000038146973
10 0.839999914169312
9 0.769999980926514
8 0.855000019073486
7 0.795000076293945
6 0.819999933242798
5 0.809999942779541
4 0.879999995231628
3 0.865000009536743
2 0.960000038146973
1 1
};
\addplot [semithick, magenta21936255]
table {%
18 0.394999980926514
17 0.419999957084656
16 0.495000004768372
15 0.519999980926514
14 0.490000009536743
13 0.605000019073486
11 0.714999914169312
10 0.759999990463257
9 0.565000057220459
8 0.745000004768372
7 0.769999980926514
6 0.789999961853027
5 0.845000028610229
4 0.815000057220459
3 0.904999971389771
2 0.975000023841858
1 1
};
\addplot [semithick, magenta]
table {%
18 0.0149999856948853
17 0.0249999761581421
16 0.0199999809265137
14 0.0299999713897705
13 0.0549999475479126
12 0.0549999475479126
11 0.110000014305115
10 0.100000023841858
9 0.115000009536743
8 0.200000047683716
7 0.215000033378601
6 0.360000014305115
5 0.444999933242798
4 0.629999995231628
3 0.75
2 0.924999952316284
1 1
};
\legend{ , , , , , ,, 0,0.2, 0.5, 1, 2,5,10,100}
\end{axis}

\end{tikzpicture}
\begin{tikzpicture}[scale=0.8]

\definecolor{cornflowerblue109146255}{RGB}{109,146,255}
\definecolor{cornflowerblue73182255}{RGB}{73,182,255}
\definecolor{cyan}{RGB}{0,255,255}
\definecolor{darkgray176}{RGB}{176,176,176}
\definecolor{deepskyblue36219255}{RGB}{36,219,255}
\definecolor{magenta}{RGB}{255,0,255}
\definecolor{magenta21936255}{RGB}{219,36,255}
\definecolor{mediumorchid18273255}{RGB}{182,73,255}
\definecolor{mediumslateblue146109255}{RGB}{146,109,255}

\begin{axis}[
tick align=center,
tick pos=both,
title={HeH$^+$ p2 error scaled},
x grid style={darkgray176},
xlabel={Readout qubits},
xmin=0, xmax=20,
xtick style={color=black},
xtick={0,5,10,15,20},
xticklabels={
  \(\displaystyle {0}\),
  \(\displaystyle {5}\),
  \(\displaystyle {10}\),
  \(\displaystyle {15}\),
  \(\displaystyle {20}\)
},
y grid style={darkgray176},
ylabel={P\(\displaystyle _{sum\, of\, 3}\)},
ymin=-0.05, ymax=1.05,
ytick style={color=black},
ytick={-0.2,0,0.2,0.4,0.6,0.8,1,1.2},
yticklabels={
  \(\displaystyle {\ensuremath{-}0.2}\),
  \(\displaystyle {0.0}\),
  \(\displaystyle {0.2}\),
  \(\displaystyle {0.4}\),
  \(\displaystyle {0.6}\),
  \(\displaystyle {0.8}\),
  \(\displaystyle {1.0}\),
  \(\displaystyle {1.2}\)
}
]
\addplot [draw=deepskyblue36219255, fill=deepskyblue36219255, mark=x,mark size=4pt, only marks]
table{%
x  y
18 0.185
17 0.395
16 0.25
15 0.57
14 0.55
13 0.67
12 0.57
11 0.75
10 0.81
9 0.7
8 0.85
7 0.91
6 0.945
5 0.805
4 0.9
3 0.955
2 1
1 1
};
\addplot [draw=cornflowerblue73182255, fill=cornflowerblue73182255, mark=x,mark size=4pt, only marks]
table{%
x  y
18 0.16
17 0.31
16 0.24
15 0.505
14 0.5
13 0.63
12 0.56
11 0.665
10 0.775
9 0.735
8 0.845
7 0.92
6 0.93
5 0.845
4 0.915
3 0.955
2 0.995
};
\addplot [draw=cornflowerblue109146255, fill=cornflowerblue109146255, mark=x,mark size=4pt, only marks]
table{%
x  y
18 0.19
17 0.295
nan nan
15 0.455
14 0.45
13 0.555
12 0.605
11 0.64
10 0.73
9 0.67
8 0.765
7 0.835
6 0.885
nan nan
4 0.87
3 0.96
2 0.985
1 1
};
\addplot [draw=mediumslateblue146109255, fill=mediumslateblue146109255, mark=x,mark size=4pt, only marks]
table{%
x  y
18 0.085
17 0.2
16 0.175
15 0.265
14 0.325
13 0.405
12 0.42
11 0.535
10 0.615
9 0.57
8 0.74
7 0.785
6 0.87
5 0.745
4 0.88
3 0.965
2 0.985
};
\addplot [draw=mediumorchid18273255, fill=mediumorchid18273255, mark=x,mark size=4pt, only marks]
table{%
x  y
17 0.085
16 0.095
15 0.115
14 0.26
13 0.165
12 0.255
11 0.37
10 0.39
9 0.39
8 0.495
7 0.615
6 0.675
5 0.705
4 0.745
3 0.885
2 0.975
1 1
};
\addplot [draw=magenta21936255, fill=magenta21936255, mark=x,mark size=4pt, only marks]
table{%
x  y
17 0.0249999999999999
nan nan
15 0.05
14 0.0600000000000001
13 0.0700000000000001
12 0.1
11 0.13
10 0.155
9 0.18
8 0.28
7 0.335
6 0.455
5 0.51
4 0.62
3 0.875
2 0.98
1 1
};
\addplot [draw=magenta, fill=magenta, mark=x,mark size=4pt, only marks]
table{%
x  y
3 0.495
2 0.805
1 1
};
\addplot [semithick, cyan]
table {%
18 0.174999952316284
17 0.399999976158142
16 0.305000066757202
15 0.559999942779541
14 0.589999914169312
13 0.625
12 0.644999980926514
11 0.795000076293945
10 0.829999923706055
9 0.769999980926514
8 0.889999985694885
7 0.964999914169312
6 0.964999914169312
5 0.865000009536743
4 0.944999933242798
3 0.970000028610229
2 0.990000009536743
1 1
};
\addplot [semithick, deepskyblue36219255]
table {%
18 0.184999942779541
17 0.394999980926514
16 0.25
15 0.569999933242798
14 0.549999952316284
13 0.670000076293945
12 0.569999933242798
11 0.75
10 0.809999942779541
9 0.700000047683716
8 0.850000023841858
7 0.910000085830688
6 0.944999933242798
5 0.805000066757202
4 0.899999976158142
3 0.954999923706055
2 1
1 1
};
\addplot [semithick, cornflowerblue73182255]
table {%
18 0.159999966621399
17 0.309999942779541
16 0.240000009536743
15 0.504999995231628
14 0.5
13 0.629999995231628
12 0.559999942779541
11 0.664999961853027
10 0.774999976158142
9 0.735000014305115
8 0.845000028610229
7 0.920000076293945
6 0.930000066757202
5 0.845000028610229
4 0.914999961853027
2 0.995000004768372
1 1
};
\addplot [semithick, cornflowerblue109146255]
table {%
18 0.190000057220459
17 0.294999957084656
16 0.245000004768372
15 0.455000042915344
14 0.450000047683716
13 0.555000066757202
12 0.605000019073486
11 0.639999985694885
10 0.730000019073486
9 0.670000076293945
8 0.764999985694885
7 0.835000038146973
6 0.884999990463257
5 0.815000057220459
4 0.870000004768372
3 0.960000038146973
2 0.985000014305115
1 1
};
\addplot [semithick, mediumslateblue146109255]
table {%
18 0.0850000381469727
17 0.200000047683716
16 0.174999952316284
15 0.264999985694885
14 0.325000047683716
13 0.404999971389771
12 0.419999957084656
11 0.535000085830688
10 0.615000009536743
9 0.569999933242798
8 0.740000009536743
7 0.785000085830688
6 0.870000004768372
5 0.745000004768372
4 0.879999995231628
3 0.964999914169312
2 0.985000014305115
1 1
};
\addplot [semithick, mediumorchid18273255]
table {%
18 0.0399999618530273
17 0.0850000381469727
16 0.0950000286102295
15 0.115000009536743
14 0.259999990463257
13 0.164999961853027
12 0.254999995231628
11 0.370000004768372
10 0.389999985694885
9 0.389999985694885
8 0.495000004768372
7 0.615000009536743
6 0.674999952316284
5 0.704999923706055
4 0.745000004768372
3 0.884999990463257
2 0.975000023841858
1 1
};
\addplot [semithick, magenta21936255]
table {%
18 0.0149999856948853
17 0.0249999761581421
16 0.0299999713897705
15 0.0499999523162842
13 0.0700000524520874
11 0.129999995231628
9 0.180000066757202
8 0.279999971389771
7 0.335000038146973
6 0.455000042915344
5 0.509999990463257
4 0.620000004768372
3 0.875
2 0.980000019073486
1 1
};
\addplot [semithick, magenta]
table {%
18 0.0149999856948853
17 0.0199999809265137
14 0.0199999809265137
13 0.0249999761581421
12 0.0349999666213989
11 0.0299999713897705
10 0.0299999713897705
8 0.059999942779541
7 0.0700000524520874
6 0.129999995231628
5 0.184999942779541
4 0.375
3 0.495000004768372
2 0.805000066757202
1 1
};
\end{axis}

\end{tikzpicture}
    \caption{Probability to read the three most likely phases for between 1 and 18 readout ancillae with the H2 ion-trap emulator two-qubit gate error rate (p2) scaled by between 0 (blue, top) and 100 (pink, bottom). `$\times$' marks denote results where the highest probability phase of the noisy result matches the highest probability phase of the memory error-free result for 200 shots. }
    \label{p2_scaled_p3max}
\end{figure}
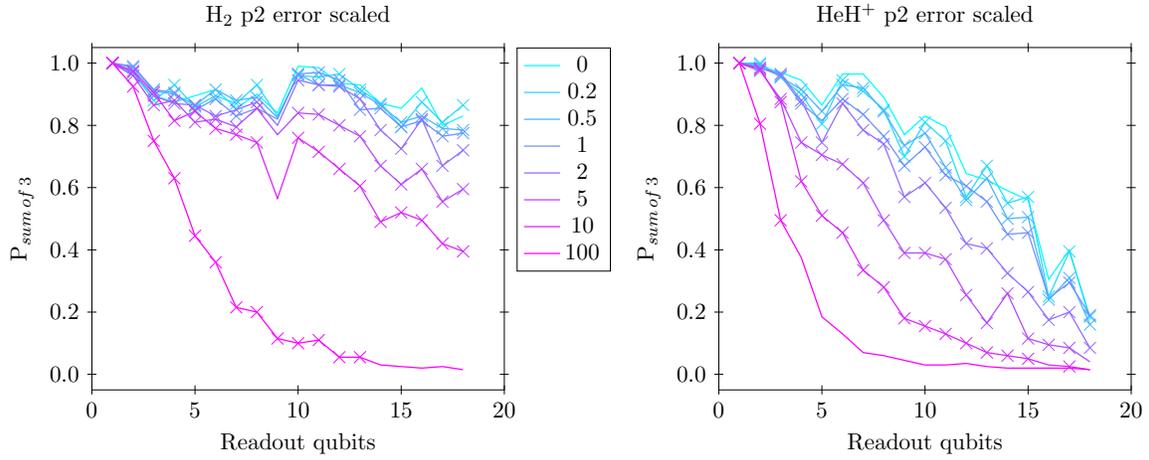

\clearpage
\bibliography{HWQPE}
\bibliographystyle{unsrtnat}

\end{document}